\documentclass[journal]{IEEEtran}      

\usepackage{graphicx}      
\usepackage{times} 
\usepackage{graphicx}  
\usepackage{caption}
\usepackage{subcaption}

\usepackage{array}
\usepackage{tikz}
\usepackage{latexsym,theorem}
\usepackage{amsmath,amsfonts,amssymb,amsbsy}
\usepackage{verbatim}
\usepackage{mathrsfs}

\usepackage{epstopdf}
\usepackage{color}
\usepackage{algorithm}
\usepackage{mathtools}
\usepackage[noend]{algpseudocode}
\DeclarePairedDelimiter{\ceil}{\lceil}{\rceil}

\newcommand{\Cset}{\mathbb{C}}
\newcommand{\Rset}{\mathbb{R}}

\newcommand{\Sset}{\mathbb{S}}
\newcommand{\Xset}{\mathbb{X}}
\newcommand{\Zset}{\mathbb{Z}}
\newcommand{\Wset}{\mathbb{W}}

\newcommand{\cP}{\mathcal{P}}

\newcommand{\cS}{\mathcal{S}}

\newcommand{\cN}{\mathcal{N}}
\newcommand{\cK}{\mathcal{K}}

\newcommand{\cB}{\mathcal{B}}
\newcommand{\cA}{\mathcal{A}}
\newcommand{\cL}{\mathcal{L}}

\newcommand{\cI}{\mathcal{I}}

\newcommand{\C}[1]{\mathbb{#1}}
\newcommand{\R}{\C{R}}
\newcommand{\N}{\C{N}}

{\theorembodyfont{\slshape}\newtheorem{theorem}{Theorem}[section]}
{\theorembodyfont{\slshape}\newtheorem{proposition}[theorem]{Proposition}}
{\theorembodyfont{\slshape}}
{\theorembodyfont{\slshape}\newtheorem{fact}[theorem]{Fact}}
{\theorembodyfont{\slshape}}
{\theorembodyfont{\slshape}\newtheorem{corollary}[theorem]{Corollary}}
{\theorembodyfont{\upshape}\newtheorem{definition}[theorem]{Definition}}
{\theorembodyfont{\upshape}\newtheorem{problem}[theorem]{Problem}}
{\theorembodyfont{\upshape}\newtheorem{remark}[theorem]{Remark}}
{\theorembodyfont{\upshape}\newtheorem{assumption}[theorem]{Assumption}}
{\theorembodyfont{\upshape}\newtheorem{motexample}{Example}}
{\theorembodyfont{\upshape}}

\title{\LARGE \bf
Sampling--based verification of Lyapunov's inequality \\ for piecewise continuous nonlinear systems
}

\author{Ruxandra Bobiti, Mircea Lazar%
\thanks{Ruxandra Bobiti and Mircea Lazar are with the Department of Electrical Engineering, Eindhoven University of Technology, The Netherlands, E-mails: r.v.bobiti@tue.nl, m.lazar@tue.nl.}}

\begin{document}

\maketitle
\thispagestyle{empty}
\pagestyle{empty}

\begin{abstract}

This paper considers a sampling--based approach to stability verification for piecewise continuous nonlinear systems via Lyapunov functions. Depending on the system dynamics, the candidate Lyapunov function and the set of initial states of interest, one generally needs to handle large, possibly non--convex or non--feasible optimization problems. To avoid such problems, we propose a constructive and systematically applicable sampling--based method to Lyapunov's inequality verification. This approach proposes verification of the decrease condition for a candidate Lyapunov function on a finite sampling of a bounded set of initial conditions and then it extends the validity of the Lyapunov function to an infinite set of initial conditions by automatically exploiting continuity properties. This result is based on multi--resolution sampling, to perform efficient state--space exploration. Using hyper--rectangles as basic sampling blocks, to account for different constraint scales on different states, further reduces the amount of samples to be verified. Moreover, the verification is decentralized in the sampling points, which makes the method scalable. The proposed methodology is illustrated through examples.

\end{abstract}

\section{Introduction}

Sampling--based analysis is an emerging methodology in the domain of nonlinear hybrid systems analysis, motivated by real-time applications and the curse of dimensionality. Typical solutions involve a deterministic or randomized approach to sampling--based analysis. In this paper, we aim at providing a deterministic framework for sampling--based verification for piecewise continuous nonlinear systems.

Most commonly, sampling approaches have been used for finite--time reachability analysis of continuous--time systems, see, e.g., \cite{dang2000verification}, \cite{girard2005reachability}, \cite{althoff2008verification}, \cite{althoff2008reachability}, or \cite{fan2015bounded}, which uses discrepancy functions for bounded--time safety verification in a simulation--based framework. For a sampling--based infinite--time reachability, i.e., safety analysis, invariance, see, e.g., \cite{kapinski2013sets}. A method similar to sampling, namely, cell--mapping \cite{van1994cell}, \cite{castillo2012cell}, uses a partitioning of the state space in cells, to discover complex attractors. However, for formal guarantees it relies on optimization or non--deterministic tools.

In what concerns the stability analysis of hybrid nonlinear systems, typically a Lyapunov function is constructed, and
its largest viable level set is computed to estimate the domain of attraction (DOA) of an equilibrium of interest \cite{khalil2002nonlinear}, \cite{vidyasagar2002nonlinear}. Most methods rely on the following common approach: verify
the decrease condition for a candidate Lyapunov function and a candidate subset of $\R^n$, which can be
a bounded or unbounded set, an infinite or finite set of states (e.g., generated by simulations
\cite{kapinski2014simulation} or state--space sampling \cite{van1994cell}, \cite{kapinski2013sets}). Depending on the system dynamics, the candidate Lyapunov function and the
set of initial states of interest, one generally needs to solve a convex or non--convex optimization problem.
The corresponding optimization problem does not scale well with the state--space dimension and in the non--convex
case, attaining a global optimum for a large set of initial states is difficult.

Sampling based approaches to computing Lyapunov functions have been developed, e.g., by \cite{topcu2008local} and \cite{kapinski2014simulation}. The work presented in \cite{topcu2008local} selects samples as starting points to generate simulation traces which are used to obtain local candidate polynomial Lyapunov functions for dynamical systems with polynomial vector fields. In \cite{topcu2008local}, simulations allow for converting a set of computationally expensive bilinear matrix inequalities into linear matrix inequalities, which are more tractable. This idea is extended in \cite{kapinski2014simulation} by a procedure to improve iteratively the quality of the candidate Lyapunov function. The procedure relies on a falsification tool in the form of a global optimizer which generates a series of successively improved intermediate Lyapunov functions. The Lyapunov function found by the simulation--based iterative technique is validated formally through queries in Satisfiability Modulo Theories (SMT) solvers such as dReal \cite{gao2012delta}, z3 \cite{demoura2008z3}, MetiTarski \cite{akbarpour2010metitarski}. 

Complementary, in \cite{berkenkamp2016safe}, a Lyapunov function is chosen based on a limited knowledge of the system and samples are generated in order to verify the Lyapunov function with high accuracy and expand the DOA of the true system  via experiments. In \cite{najafi2016fast}, a fast sampling--based method for estimating the DOA in real--time was proposed, though, without formal guarantees.

In this paper, a methodology is developed for finding a Lyapunov function and verifying its validity for piecewise continuous nonlinear systems. The essence of the approach is to construct a candidate Lyapunov function via the converse result in \cite{geiselhart2013alternative}, and to verify the decrease condition for the candidate Lyapunov function on a finite sampling of a bounded set of
initial conditions and then to extend the validity of the Lyapunov function to an infinite set of initial conditions by exploiting
continuity properties. The verification for the points in the finite set of samples is independently performed and therefore the methodology is spatially decentralized. This feature makes this approach applicable to sets which do not fully satisfy the decrease condition of the Lyapunov function and would deem an optimization problem unfeasible. Moreover, building the candidate Lyapunov function via the converse result in \cite{geiselhart2013alternative} allows for "freely" choosing a candidate function and embedding the verification problem in the construction of the Lyapunov function by iteratively increasing the decrease step of the candidate function if the current step does not verify the decrease condition.

The main contributions of this paper are the following. A sampling--based verification framework is developed for an inequality of the type $F(x)\leq (<)0$, for all $x\in\cS$, where $F:\Rset^n\rightarrow \Rset$ may be piecewise continuous and $S\subset\R^n$ is a compact set. Particularly, in this paper, verification of the Lyapunov's inequality is addressed, based on hyper--rectangle sampling of the state space. The methodology presented here is applicable to discrete--time systems. To verify Lyapunov's inequality for continuous--time systems, the same methodology is applied for the discretized system to find a candidate Lyapunov function, and additionally the Lyapunov's inequality is finally validated for the continuous--time system. Lastly, using the same sampling--based verification tools, we present a method for computing the level set of the Lyapunov function computed previously on a possibly non--convex set.


The remainder of this paper is organized as follows. In Section~\ref{prelim}, preliminary notations and instrumental stability results are introduced. The main contributions of the paper are presented in Section~\ref{mainresult}, which consists of a theoretical result for decentralized sampling--based verification, with implementation details. Section~\ref{stability} adapts the derived methodology for stability analysis of discrete--time and continuous--time systems and computation of level sets of Lyapunov functions. Illustrative examples are provided in Section~\ref{examples}, and Section~\ref{conclusions} concludes the paper.

\section{Preliminaries}
\label{prelim}

\subsection{Basic notation and definitions}

Let $\mathbb{R}$, $\mathbb{R}_+$, $\mathbb{Z}$ and $\mathbb{Z}_+$ denote the field of real numbers, the set of non--negative reals, the set of integers and the set of non--negative integers, respectively. For every $c\in\Rset$ and $\Pi\subseteq\Rset$, define $\Pi_{\geq c}:=\{k\in\Pi\mid k\geq c \}$ and similarly $\Pi_{\leq c}$. Let $int(\mathbb{S})$ denote the interior of a set $\mathbb{S}$. Let $\mathbb{S}^h:=\mathbb{S}\times\ldots\times\mathbb{S}$ for any $h\in\mathbb{Z}_{\geq1}$ denote the $h$--times Cartesian--product of $\mathbb{S}\subseteq\Rset^n$. Denote $\circ$ the operator of maps composition, i.e., for two arbitrary maps $\alpha_1:\mathbb{D}_{1}\rightarrow\mathbb{C}_{1}$, and $\alpha_2:\mathbb{D}_{2}\rightarrow\mathbb{C}_{2}$, with $\mathbb{C}_{2}\subseteq \mathbb{D}_{1}$,  $\alpha_1\circ\alpha_2(x)=\alpha_1(\alpha_2(x))$, for all $x\in\mathbb{D}_{2}$. Let $\alpha^h:=\alpha\circ \ldots \circ\alpha$ for any $h\in\mathbb{Z}_{\geq1}$ denote the $h$--times map composition of $\alpha:\Cset\rightarrow\Cset$. Define the identity function by $id:\Sset\rightarrow\Sset$ such that for any $x\in\Sset$, $id(x)=x$. The operator $\oplus$ denotes the Minkowski sum, i.e., $A\oplus B:=\{a+b: a\in A, b\in B\}$.  Denote by $\bar{A}$ the closure of the set $A$. A set $\cS\subset\Rset^n$ is called proper if it is non--empty, compact and $0\in int(\cS)$. Given a proper set $\cS\subset\Rset^n$, for any $\cN(0)$, i.e., a neighborhood of 0, the set $\cA:=\overline{\cS\setminus\cN(0)}$ is an annulus of $\cS$.

For a vector $x\in\Rset^n$, the symbol $\|x\|$ is used to denote an arbitrary $p$--norm; it will be made clear when a specific norm is considered. The absolute value of the vector $x$, i.e., $|x|$, is the vector of the absolute values of the elements in $x$. For a scalar $x\in\Rset$, denote by $\ceil{x}$ the smallest integer number larger than $x$.

  \begin{figure}[!htpb]
      \centering
      \includegraphics[width=1\columnwidth]{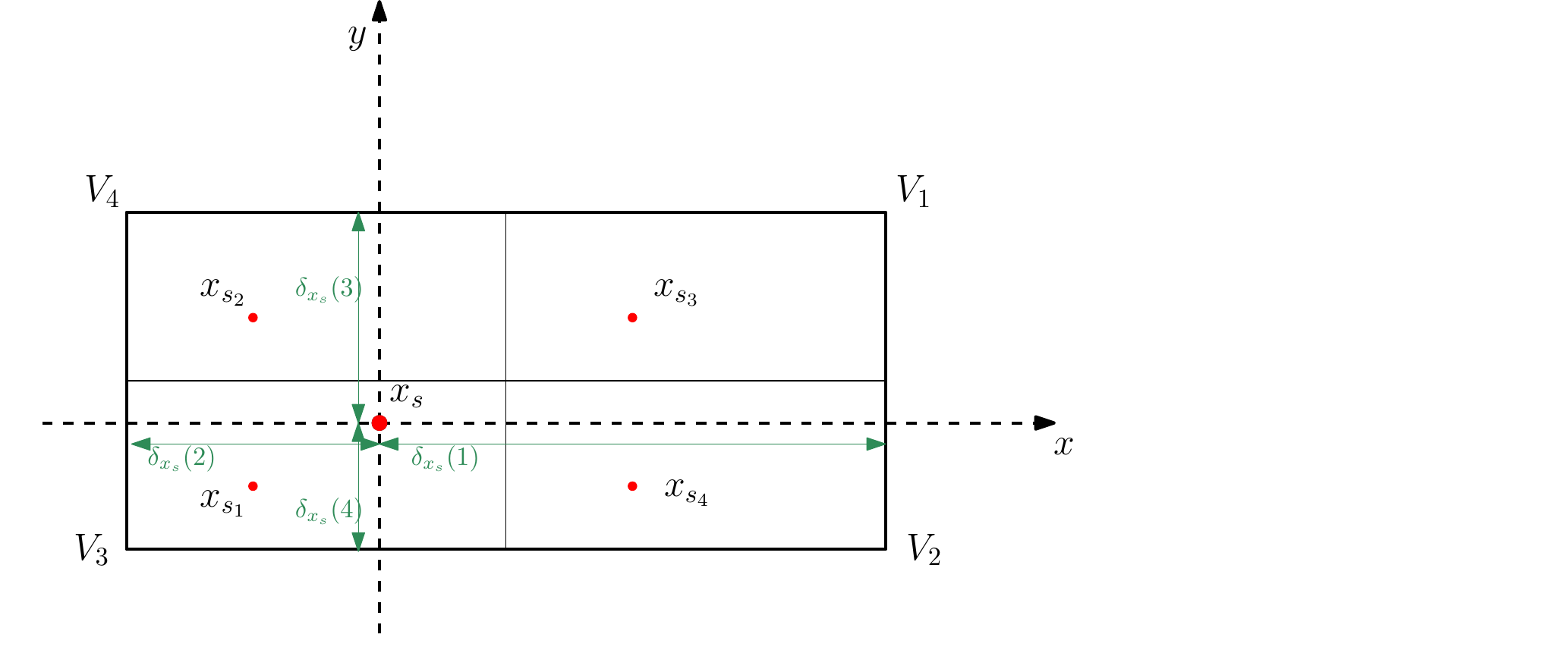}
      \caption{2D hyper--rectangle refinement illustration.}
      \label{hyper}
\end{figure}

A hyper--rectangle of dimension $n$, centered in $x_s\in\Rset^n$, see Fig.~\ref{hyper} for $n=2$, is described as follows. If $V_j\in\Rset^n$ with $j\in\Zset_{[1, 2^n]}$ are the vertices of the hyper--rectangle, then a vector $\delta_{x_s}\in\Rset^{2n}$ can be computed as follows:
  $$\delta_{x_s}(2i-1)=\max_{j\in\Zset_{[1, 2^n]}}\{V_j(i)-x_s(i)\},$$
  $$\delta_{x_s}(2i)=\min_{j\in\Zset_{[1, 2^n]}}\{V_j(i)-x_s(i)\},$$
  for all $i\in\Zset_{[1, n]}$.

  The hyper--rectangle has the hyper--plane representation $\cB_{\delta_{x_s}}(x_s):=\{\xi\in\Rset^n: \max_{i\in\Zset_{[1, 2n]}}[P_{x_s}]_{i:}(\xi-x_s)\leq 1 \},$ where
\[ P_{x_s}:=\left( \begin{array}{cccc}
\frac{1}{\delta_{x_s}(1)} & 0 & \ldots & 0 \\
\frac{1}{-\delta_{x_s}(2)} & 0 & \ldots & 0 \\
\vdots & \vdots & \ddots & \vdots \\
0 & 0 & \ldots & \frac{1}{\delta_{x_s}(2n-1)} \\
0 & 0 & \ldots & \frac{1}{-\delta_{x_s}(2n)} \end{array} \right).\]
   Notice that $\cB_{\delta_{x_s}}(x_s)$ is represented through a gauge function inequality. If the hyper--rectangle is a hyper--cube, as the basic sampling unit in \cite{bobiti2015seattle}  and \cite{bobiti2016ecc}, then $\delta_{x_s}$ can be reduced to a scalar, and the sampling unit is represented through a norm inequality:
  $\cB_{\delta_{x_s}}(x):=\{\xi\in\Rset^n: \|\xi-x_s\|\leq \delta_{x_s} \}$, i.e., a symmetric gauge function inequality. Let $\cB_{\delta}:=\cB_{\delta}(0)$.

A function $\alpha:\mathbb{R}_+\rightarrow\mathbb{R}_+$ is said to belong to class $\cK$, i.e., $\alpha\in\cK$, if it is continuous, strictly increasing and $\alpha(0)=0$. Furthermore, $\alpha\in\cK_\infty$ if $\alpha\in\cK$ and $\lim_{s\rightarrow\infty}\alpha(s)=\infty$. The function $\beta:\Rset_+\times\Rset_+\rightarrow\Rset_+$ is said to belong to class $\cK\cL$, i.e., $\beta\in\cK\cL$, if for each fixed $s\in\Rset_+$, $\beta(\cdot,s)\in\cK$ and for each fixed $r\in\Rset_+$, $\beta(r,\cdot)$ is decreasing and $\lim_{s\rightarrow\infty}\beta(r,s)=0$.

\begin{definition}\label{definition:kcont}
Let $\cS\subseteq\Rset^n$. Then, a map $G:\cS\rightarrow \cS$ is called $\cK$--continuous in $\cS$ if there exists a function $\sigma\in\cK$ such that
\begin{align}\label{eq:keq}
\|G(x)-G(y)\|\leq \sigma(\|x-y\|),\quad \forall (x, y)\in\cS\times \cS.
\end{align}
\end{definition}

We call $\sigma$ the continuity function of the map $G$. If $\sigma(s)=a s$ with $a\in\Rset_+$, then $\cK$--continuity recovers Lipschitz continuity.

\subsection{Stability analysis tools}

Consider the autonomous nonlinear system in discrete--time
\begin{align}\label{eq:nondist}
x_{k+1}=G(x_k),\quad k\in\Zset_+,
\end{align}
and in continuous--time
\begin{align}\label{eq:nondist_ct}
\dot{x}=G_c(x),
\end{align}
where $x_k \in \cS$ (resp. $x \in \cS$) is the state, $\cS$ is a compact set with $0\in int(\cS)$, and $G:\Rset^n\rightarrow \Rset^n$, $G_c:\Rset^n\rightarrow \Rset^n$ are piecewise continuous nonlinear functions, and $G_c$ is locally Lipschitz. A point $x^*\in\cS$ is an equilibrium point of system \eqref{eq:nondist} if $G(x^*)=x^*$, and of system \eqref{eq:nondist_ct} if $G_c(x^*)=0$. We assume $G(0)=0$ and $G_c(0)=0$. The domain of attraction (DOA) of the origin is the set of all initial states, from which the state trajectories asymptotically converge to the origin. Denote the solution of \eqref{eq:nondist_ct} with initial state $x(0)$ at time $t=0$ by $x(t)$ for any $t\in\Rset_{\geq 0}$. Assume that $x(t)$ exists and it is unique for all $t\in\Rset_{\geq 0}$. For system \eqref{eq:nondist}, 
define the one--step reachable set from $\cS$ as $Reach(\cS):=\cup_{\xi\in\cS}G(\xi)$. 

\begin{definition}\label{definition:klstabdef}
The system \eqref{eq:nondist} (resp. \eqref{eq:nondist_ct}) is called $\cK\cL$--stable on $\cS$ if there exists a $\cK\cL$ function $\beta:\Rset_+\times\Rset_+\rightarrow \Rset_+$ such that $\|x_{k+1}\|\leq \beta(\|x_0\|, k)$ for all $(x_0,k)\in\cS\times\Zset_+$ (resp. $\|x(t)\|\leq \beta(\|x_0\|, t)$ for all $(x_0,t)\in\cS\times\Rset_+$).
\end{definition}

\begin{proposition}\label{thm:2suf_compact} \cite{bobiti2014sinaia}
Let $\Wset$ be a compact set with $0\in int(\Wset)$, which is invariant with respect to the dynamics \eqref{eq:nondist} (resp.  \eqref{eq:nondist_ct}). Let $\alpha_1, \alpha_2\in\cK_{\infty}$. Suppose that the map $G$ corresponding to the dynamics \eqref{eq:nondist} is $\cK$--bounded on $\Xset$ and there exists a function $W:\Rset^n\rightarrow\Rset_+$ such that
\begin{subequations}
\begin{align}
\alpha_1(\|x\|)\leq W(x)\leq \alpha_2(\|x\|),\quad\forall x\in \Wset,\label{eq:thm2acompact}
\end{align}
and there exists an $M\in\Zset_{\geq 1}$ and a corresponding $\rho\in\cK$ with $\rho<id$ such that
\begin{align}
W(G(x))\leq\rho (W(x)),\quad \forall x\in\Wset,\label{eq:thm2bcompact}
\end{align}
for \eqref{eq:nondist} (resp.
\begin{align}
\dot{W}(x)< 0,\quad \forall x\in\Wset\backslash\{0\}.\label{eq:thm2bcompact_ct}
\end{align}
for \eqref{eq:nondist_ct}).
\end{subequations}
Then, $W$ is a Lyapunov function on $\Wset$ and system~\eqref{eq:nondist} (resp. \eqref{eq:nondist_ct}) is $\cK\cL$-stable in $\Wset$. 
\end{proposition}

Let $V:\Rset^n\rightarrow\Rset_+$ satisfying
 \begin{equation}\label{eq:fslf}
 V(x)=\eta(\|x\|)
 \end{equation}
 for some $\eta\in\cK_{\infty}$ and some norm $\|\cdot\|$. If there exists an $M\in\Zset_{\geq 1}$ and a corresponding $\rho\in\cK$ with $\rho<id$ such that
 \begin{align}
V(G^M(x))\leq\rho (V(x)),\quad \forall x\in\Wset.\label{eq:ftlf_ineq}
\end{align}
then, by the converse theorem in \cite[Theorem 20]{geiselhart2013alternative}, adapted in \cite{bobiti2014sinaia} for compact sets, the function $W:\Rset^n\rightarrow\Rset_+$ which satisfies
\begin{align}
\label{eq:lf}
W(x)=\sum_{j=0}^{M-1}V(G^j(x)),
\end{align}
is a Lyapunov function.

These definitions are instrumental for verifying stability with the sampling--based verification framework proposed in the next section.

\section{Sampling--based verification}
\label{mainresult}

This section considers a general sampling--based verification problem, as formulated below.

Given a finite sampling $\cS_s$ of a compact set $\cS\subset\Rset^n$, for all $x\in\cS$, there exists at least one pair $(x_s, \delta_{x_s})\in \cS_s\times \Rset^{2n}$ s.t. $\|x-x_s\|\leq\max(|\delta_{x_s}|)$
and $\cS\subseteq \cup_{x_s\in \cS_s}\cB_{\delta_{x_s}}(x_s)$. When this property holds for a specific set $\Delta:=\{\delta_{x_s}: (x_s, \delta_{x_s})\in\cS_s\times\Rset^{2n} \}$, we call the set $\cS_s$ a $\Delta$--sampling of $\cS$.

\begin{problem}\label{prob}
 Given a sampling $\cS_s$ of a compact set $\cS$ and a real valued, piecewise continuous function $F\ : \cS\rightarrow\R$, construct a function $\gamma:\Rset_+\times \cS_s\rightarrow \Rset_+$ such that if $F(x_s)\leq -\gamma(\delta_{x_s}, x_s)<0$ for all $x_{s}\in\cS_s$, then $F(x)\leq 0$ for all $x\in\cS$.
\end{problem}

In what follows we develop a fully decentralized solution to Problem~\ref{prob}.

\subsection{Decentralized sampling--based verification}

Let the sets $\cS_i$ with $i\in\cI:=\{1,\ldots,N\}$ for some $N\in\N$ satisfy $\cup_{i\in\cI}\cS_i=\cS$, i.e., the sets $\cS_i$ define a partition of the compact set $\cS$. Given a $\Delta$--sampling $\cS_s$ of $\cS$, define
$$\cS_s^i:=\{x_s: \cB_{\delta_{x_s}}(x_s)\cap \cS_i \neq \varnothing\},$$
$$\Delta_i:=\{\delta_{x_s}: (x_s, \delta_{x_s})\in\cS_s^i\times \Rset^{2n}\},$$
$$I_{x_s}:=\{i\in\cI: \cB_{\delta_{x_s}}(x_s)\cap \cS_i\neq \varnothing\},$$
$$\delta_i:=\max_{\delta_{x_s}\in\Delta_i}|\delta_{x_s}|.$$
Recall $|\delta_{x_s}|$ is the vector of absolute values of $\delta_{x_s}$. Let $F_i\ :\cB_{\delta_i}\oplus \cS_i\rightarrow\R$ be real valued continuous functions for all $i\in\cI$. Note that $\Delta=\cup_{i\in\cI}\Delta_i$. The following assumptions are instrumental in what follows.
\begin{assumption}\label{exists11}
We assume that the following property holds:
\begin{equation}
\label{eq3.1}
\forall x\in\cS, \quad \exists_1 i\in\cI \ : \ F(x)=F_i(x),
\end{equation}
where $\exists_1$ denotes that there exists a unique element.
\end{assumption}

\begin{assumption}\label{ficont1}
We assume that, for any $x_s\in\cS_s$, there exist $a_{x_s}^i, b_{x_s}^i\in\Rset$ such that:
\begin{equation}
\label{eq3.222}
|F_i(x)-F_i(x_s)|\leq a_{x_s}^i\|x-x_s\|+b_{x_s}^i,
\end{equation}
for all $i\in I_{x_s}$ and $x\in\cB_{\delta_{x_s}}(x_s)$.
\end{assumption}

If we write $a_{x_s}:=\max_{i\in I_{x_s}} a_{x_s}^i$ and $b_{x_s}:=\max_{i\in I_{x_s}} b_{x_s}^i$, then \eqref{eq3.222} implies
\begin{equation}
\label{eq3.2}
|F_i(x)-F_i(x_s)|\leq a_{x_s}\|x-x_s\|+b_{x_s} ,
\end{equation}
for all $i\in I_{x_s}$ and $x\in\cB_{\delta_{x_s}}(x_s)$.

Assumption~\ref{ficont1} does not imply that $F$ is continuous on $\cS$. For example, $F$ could be constructed by switching among the different continuous $F_i$ functions within the partition of the set $\cS$, see Example~\ref{example}. Next, define the set--valued regularization map $\overline{F}\ : \ \cS\rightrightarrows\R$ as \[\overline{F}(x):=\bigcap_{\rho>0}\overline{\bigcup_{\epsilon\in\cB_{\rho}}F(x+\epsilon)}.\] For all $x\in\cS$ define the index set:
\[I(x):=\{i\in\cI\ : \ F_i(x)\in\overline{F}(x)\}.\]
Furthermore, define the real--valued function $\varepsilon \ : \ \R^n \rightarrow\R_+$,
\[\varepsilon(x):=\max\{|F_i(x)-F_j(x)|\ : \ (i,j)\in I(x)\times I(x)\}.\]
Observe that $\varepsilon$ yields the maximum absolute jump that can occur in the function $F$ at a point $x$, due to discontinuity. If $F$ is continuous at $x$ clearly $I(x)$ is a singleton and consequently $\varepsilon(x)=0$. As such, if $F$ is continuous on $\cS$, then $\varepsilon(x)=0$ for all $x\in\cS$.

\begin{motexample}\label{example}
To illustrate the notions introduced so far, consider the system \cite[Example 4]{luk2015domain}:

\[
x^+: = G(x)=
  \begin{cases}
   G_1(x) & \text{if } x\in S_1 \\
   G_2(x) & \text{if } x\in S_2,
  \end{cases}
\]
where
$$G_1(x)=\left[\begin{matrix} 0.5x_1 & -0.8x_2-x_1^2 \end{matrix}\right]^T,$$
$$G_2(x)=\left[\begin{matrix} 0.5x_1+x_1x_2 & -0.8x_2 \end{matrix}\right]^T,$$
$$\cS=\{x\in\Rset^2: \|x\|_{\infty}\leq 1.5\}$$
$$S_1:=\{x\in\Rset^2: x_2\geq 0\}\cap\cS,\text{ } S_2:=\{x\in\Rset^2: x_2< 0\}\cap\cS.$$
Notice that the system is discontinuous on the first axis. Consider a sampling point $x_s=\left[\begin{matrix} 1 & 0 \end{matrix}\right]^T$. We want to compute $\varepsilon(x_s)$ for the function $F:\Rset^2\rightarrow\Rset$ defined by $F(x)=V(G^3(x))-\rho V(x)$, where $V:\Rset^2\rightarrow\Rset_+$ is defined by $V(x)=x^Tx$ and $\rho\in\Rset_{[0,1]}$. Notice that if $F(x)\leq 0$ for a specific $x\in\Rset^n$, then \eqref{eq:ftlf_ineq} holds for that $x$, with $M=3$. Furthermore, $I(x_s)=\{1, 2\}.$

To compute $F_1(x_s)$ evaluate $G_1(x_s)=\left[\begin{matrix} 0.5 & -1 \end{matrix}\right]^T$. Since the second element in this vector is less than 0, then $G^2(x_s)=G_2(G_1(x_s))=\left[\begin{matrix} -0.25 & 0.8 \end{matrix}\right]^T$, and similarly, $G^3(x_s)=G_1(G_2(G_1(x_s)))=\left[\begin{matrix} -0.125 & -0.7025 \end{matrix}\right]^T$. Therefore, $F_1(x_s)=V(G_1(G_2(G_1(x_s))))-\rho V(x_s)=0.5091-\rho$. Similarly, for computing $F_2(x_s)$ we evaluate $G_2(x_s)=\left[\begin{matrix} 0.5 & 0 \end{matrix}\right]^T$, $G_1(G_2(x_s))=\left[\begin{matrix} 0.25 & -0.25 \end{matrix}\right]^T$ and $G^3(x_s)=G_2(G_1(G_2(x_s)))=\left[\begin{matrix} 0.0625 & 0.2 \end{matrix}\right]^T$, which gives $F_2(x_s)=V(G_2(G_1(G_2(x_s))))-\rho V(x_s)=0.0439-\rho$. Therefore, $\varepsilon(x_s)=|F_1(x_s)-F_2(x_s)|=0.4652$, which illustrates the discontinuity of $F(x)$.

\end{motexample}

Under the above assumptions and definitions, let us state the main sampling verification theorem.

\begin{theorem}\label{thm:general}
Suppose Assumption~\ref{exists11} and Assumption~\ref{ficont1} hold. Let $\cS_s$ be a $\Delta$--sampling of the set $\cS$ and let $F:\cS\rightarrow \Rset$ and the associated functions $F_i:\cB_{\delta}\oplus \cS_i\rightarrow\R$ be given. If for all $x_s\in \cS_s$ it holds that
\begin{align}\label{eq:toverify_hybrid}
F(x_s) & \leq -\bar{\gamma}(\max |\delta_{x_s}|, x_s) \nonumber \\
\mbox{ (resp. } F(x_s) & < -\bar{\gamma}(\max |\delta_{x_s}|, x_s)\mbox{ ),}
\end{align}
where $\bar{\gamma}:\Rset_+\times \cS_s \rightarrow \Rset_+$ and $\bar{\gamma}(\xi, x_s):=a_{x_s}\xi+b_{x_s}+\varepsilon(x_s)$, then $F(x)\leq 0$ (resp. $F(x)< 0$) holds for all $x\in \cS$.
\end{theorem}

\begin{IEEEproof} Assume $F(x_s)\leq -\bar{\gamma}(\max |\delta_{x_s}|, x_s)$ (resp. $F(x_s)< -\bar{\gamma}(\max |\delta_{x_s}|, x_s)$) holds for all  $x_s\in \cS_s$, but there exists a  point $x\in \cS$ such that $F(x)>0$ (resp. $F(x)\geq 0$). Take any point $x_s\in \cS_s$ such that $\|x-x_s\|\leq \max |\delta_{x_s}|$. Observe that such a point always exists, by the definition of a $\Delta$--sampling of a set. By Assumption~\ref{ficont1} it follows that \eqref{eq3.2} holds for all $i\in I_{x_s}$.

Furthermore, let $i\in\cI$ be such that $F(x)=F_i(x)$ and let $j\in\cI$ be such that $F(x_s)=F_j(x_s)$.

Then, by the triangle inequality and \eqref{eq3.2} it follows that
\begin{align}
\label{eq:eqqq1}
|F(x)-F(x_s)|&=|F_i(x)-F_j(x_s)|\notag\\
&=|F_i(x)-F_i(x_s)+F_i(x_s)-F_j(x_s)|\notag\\
&\leq |F_i(x)-F_i(x_s)|+|F_i(x_s)-F_j(x_s)|\notag\\
&\leq a_{x_s}\|x-x_s\|+b_{x_s}+\varepsilon(x_s)\notag\\
&=\bar{\gamma}(\|x-x_s\|, x_s)\notag\\
&\leq a_{x_s}\max |\delta_{x_s}|+b_{x_s}+\varepsilon(x_s) \notag\\
& =\bar{\gamma}(\max |\delta_{x_s}|, x_s).
\end{align}

Since $F(x)>0$ (resp. $F(x)\geq 0$) for some $x\in \cS$, then
\begin{equation}\label{eq:eqq2}
-F(x)<0 \mbox{ (resp. } -F(x)\leq 0 \mbox{ ).}
\end{equation}
Also, for any $x_s\in \cS_s$ such that $\|x-x_s\|\leq \max |\delta_{x_s}|$ we have
\begin{align}\label{eq:eqq3}
F(x_s) & \leq -\bar{\gamma}(\max |\delta_{x_s}|, x_s) \nonumber \\
\mbox{ (resp. } F(x_s) & < -\bar{\gamma}(\max |\delta_{x_s}|, x_s) \mbox{ ).}
\end{align}
By summing up \eqref{eq:eqq2} and \eqref{eq:eqq3} we obtain:
\begin{equation}\label{eq:eqq4}
F(x_s)-F(x)<-\bar{\gamma}(\max |\delta_{x_s}|, x_s)<0,
\end{equation}
and therefore
\begin{equation}\label{eq:eqq5}
\vert F(x)-F(x_s)\vert>\bar{\gamma}(\max |\delta_{x_s}|, x_s).
\end{equation}
By inspecting \eqref{eq:eqqq1} and \eqref{eq:eqq5} we observe that a contradiction was reached. Hence, the hypothesis holds true.
\end{IEEEproof}

Decentralization is achieved by allowing $\bar{\gamma}$ to have different coefficients $a_{x_s}$ and $b_{x_s}$ for each sampling point $x_s$ and formulating the result in Theorem~\ref{thm:general} without any central variable, i.e., with no common variable for all $x_s \in S_s$.

\begin{remark} In the case that $F$ is continuous on $S$, then $\varepsilon(x_s)=0$ for all $x_s\in\cS_s$, and the result in \cite[Theorem III.3]{bobiti2016ecc} is recovered with $\bar{\gamma}(\max | \delta_{x_s}|, x_s)=a_{x_s}\max | \delta_{x_s}|+b_{x_s}$. In \cite{bobiti2016ecc}, however, the elements in $\delta_{x_s}$ are identical.
\end{remark}

\begin{remark} In the case that $F$ is piecewise continuous on $\cS$ and $\bar{\gamma}:\Rset_+\rightarrow\Rset_+$, $\bar{\gamma}(\|x-x_s\|) =\sigma(\|x-x_s\|)+\varepsilon(x_s)$, where $\sigma$ is the $\cK$--continuity function of $F$, Theorem~\ref{thm:general} reduces to \cite[Theorem 3.3]{bobiti2015seattle}. Therein, the conditions to be verified were $\sigma(\delta)<\gamma, F(x_s)\leq -\gamma-\varepsilon(x_s)$, with a variable $\gamma$ central to all sampling points $x_\delta\in\cS_s$, which requires the result in \cite{bobiti2015seattle} to be verified in a centralized manner. Notice that these two conditions can be rewritten as $F(x_s)<-\gamma-\varepsilon(x_s)$, which is decentralized. However, it is difficult to compute $\sigma$ for general nonlinear systems.
\end{remark}

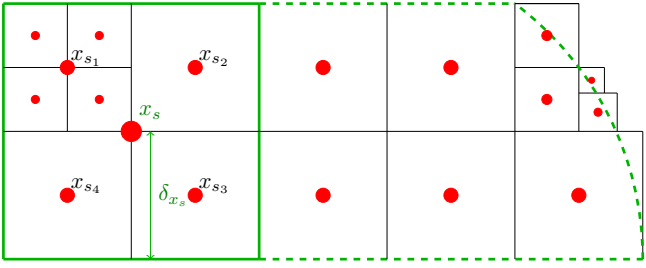
\begin{figure}[!htpb]
      \centering
      \vspace{0.2cm}
      \scalebox{0.85}{\begin{tikzpicture}
    \draw[very thick, black!30!green] (-4,0) -- (0, 0);
    \draw[very thick, black!30!green, dashed] (0, 0)-- (6,0);
    \draw[very thick, black!30!green] (0,0) -- (0,4);
    \draw (0,2) -- (4,2);
    \draw[very thick, black!30!green] (-4,4) -- (0, 4);
    \draw[very thick, black!30!green, dashed] (0, 4) -- (4,4);
    \draw[very thick, black!30!green] (-4,0) -- (-4,4);
    \draw (4,0) -- (4,4);
    \draw (2,0) -- (2,4);
    \draw (4,2) -- (6,2);
    \draw (6,0) -- (6,2);
    \draw (4,3) -- (5,3);
    \draw (5,2) -- (5,3);
    \draw (5,2.6) -- (5.6,2.6);
    \draw (5.6,2) -- (5.6,2.6);
    \draw (5,3) -- (5.4,3);
    \draw (5.4,2.6) -- (5.4,3);
    \draw (5,3) -- (5,4);
    \draw (4,4) -- (5,4);
    \draw (-4,2) -- (0,2);
    \draw (4,4) -- (5,4);
    \draw[<->, black!30!green] (-1.7,0) -- (-1.7,2) node [pos=0.5,right, black!50!green] {$\delta_{x_s}$} ;
    \draw (-2,0) -- (-2,2);
    \draw (-2,2) -- (-2,4);
    \draw (-3,2) -- (-3,4);
    \draw (-4,3) -- (-2,3);

    \draw[very thick, black!30!green, dashed] (6,0) arc (0:53.3:5cm);

    \draw[red,fill=red] (-2,2) circle (1ex);
    \draw[red,fill=red] (1,3) circle (0.7ex);
    \draw[red,fill=red] (3,3) circle (0.7ex);
    \draw[red,fill=red] (3,1) circle (0.7ex);
    \draw[red,fill=red] (5,1) circle (0.7ex);
    \draw[red,fill=red] (1,1) circle (0.7ex);
    \draw[red,fill=red] (4.5,2.5) circle (0.5ex);
    \draw[red,fill=red] (4.5,3.5) circle (0.5ex);
    \draw[red,fill=red] (5.3,2.3) circle (0.4ex);
    \draw[red,fill=red] (5.2,2.8) circle (0.3ex);

    \draw[red,fill=red] (-3.5,3.5) circle (0.4ex);
    \draw[red,fill=red] (-3.5,2.5) circle (0.4ex);
    \draw[red,fill=red] (-2.5,3.5) circle (0.4ex);
    \draw[red,fill=red] (-2.5,2.5) circle (0.4ex);

    \node[draw=none, black!50!green] at (-1.7,2.3) {$x_s$};
    \node[draw=none] at (-2.7,3.15) {$x_{s_1}$};
    \node[draw=none] at (-0.7,3.15) {$x_{s_2}$};
    \node[draw=none] at (-0.7,1.15) {$x_{s_3}$};
    \node[draw=none] at (-2.7,1.15) {$x_{s_4}$};

    \draw[red,fill=red] (-3,3) circle (0.7ex);
    \draw[red,fill=red] (-1,3) circle (0.7ex);
    \draw[red,fill=red] (-1,1) circle (0.7ex);
    \draw[red,fill=red] (-3,1) circle (0.7ex);

\end{tikzpicture} }
      \caption{Set sampling and set refinement.}
      \label{fig2}
\end{figure}

Two ingredients are required for the verification of Theorem~\ref{thm:general}. Firstly, a method for obtaining the sampling $\cS_s$ is required, which is proposed in Section~\ref{multires}. Secondly, in Section~\ref{automatic}, the procedure for computing the function $\bar{\gamma}$ is presented.

\subsection{Multi--resolution sampling tools}
\label{multires}

\begin{definition}\label{definition:refinement}
Let $x_s\in\Rset^n$, $\delta_{x_s}\in\Rset^{2n}$ be arbitrarily chosen. An $N$--refinement of $\cB_{\delta_{x_s}}(x_s)$, with $N\in\Zset_{>1}$ is a finite set $\cS_{sx_s}\subset \cB_{\delta_{x_s}}(x_s)$ s.t. for all $ x\in\cB_{\delta_{x_s}}(x_s)$, there exists at least one vector $x' \in \cS_{sx_s}$ s.t. $\|x-x'\|\leq\max(|\frac{\delta_{x_s}}{N}|)$.
\end{definition}

Notice that $\cB_{\delta_{x_s}}(x_s)\subseteq \cup_{x'\in \cS_{sx_s}}\cB_{\frac{\delta_{x_s}}{N}}(x')$.
A 2--refinement will be used throughout this paper, because it provides the minimum amount of sampling points $x'$ which refine $\cB_{\delta_{x_s}}(x_s)$ without overlay if the $\infty$--norm is used. The 2--refinement will be referred to simply as a \textit{refinement}. The refinement allows for multi--resolution sampling of the state--space. See Fig.~\ref{fig2} for an exemplification of the concept of set sampling and set refinement.

In \cite{bobiti2015seattle}  and \cite{bobiti2016ecc}, a hyper--cube was proposed as a sampling unit. This is useful when the dimension of the search set $\cS$ is similar on different axes. However, system state constraints generally present different bounds on the different axes, see for example, the example in Section~\ref{2D}. In this case, a non--uniform sampling  based on hyper--rectangles could reduce significantly the number of sampling points.

If we want to sample a hyper--rectangle $\cP$ in $n$ dimensions and $V_1, \ldots, V_{2^n}$ are the vertices of $\cP$, then by refinement of $\cP$ on $n_r<n$ dimensions, $2^{n_r}$ sampling points are obtained, and they are
$$\left\lbrace \frac{x_s+V_1}{2}, \ldots, \frac{x_s+V_{2^{n_r}}}{2}\right\rbrace,$$
with their corresponding intervals.

The sampling strategy presented so far will be illustrated through examples. However, the sampling--based verification result is independent of the sampling strategy.

\subsection{Computation of $\bar{\gamma}$ via linearization}
\label{automatic}

This subsection provides a constructive method to compute the coefficients used to define the function $\bar{\gamma}$. These computations are necessary for the verification of the statements of Theorem~\ref{thm:general}.

To verify \eqref{eq:toverify_hybrid}, for each $x_s\in\cS_s$, the following steps are required:
\begin{enumerate}
\item determine $I_{x_s}$ and $\varepsilon(x_s)$;
\item for all $i\in I_{x_s}$ verify \eqref{eq3.222}, i.e., compute $a_{x_s}^i$ and $b_{x_s}^i$;
\item \eqref{eq:toverify_hybrid} holds with $\bar{\gamma}(\max |\delta_{x_s}|, x_s):=a_{x_s}\max |\delta_{x_s}|+b_{x_s}+\varepsilon(x_s)$, where $a_{x_s}:=\max_{i\in I_{x_s}} a_{x_s}^i$ and $b_{x_s}:=\max_{i\in I_{x_s}} b_{x_s}^i$.
\end{enumerate}

In the remainder of this subsection, a constructive method is presented for the computation of $a_{x_s}^i$ and $b_{x_s}^i$, as requested in step 2). For this purpose, the following assumption is instrumental.
\begin{assumption}\label{differentiabillity}
Assume that $F_i$ is continuous and at least two times differentiable on $\cB_{\delta_{x_s}}(x_s)$ for all $x_s\in S_s$.
\end{assumption}

Notice that $\cB_{\delta_{x_s}}(x_s)$ is a convex set for all $x_s\in\cS_s$. Assumption \ref{differentiabillity} allows for Taylor series expansion and application of the Mean Value theorem in the following manner.

Denote:
  $$T(x, x_s, m):=\sum_{v=0}^m\frac{([(x-x_s)\nabla]^v F_i)(x_s)}{v!},$$
for all $x\in\cB_{\delta_{x_s}}(x_s)$, where, e.g., $\nabla F_i$ stands for the Jacobian and $\nabla^2 F_i$ is the Hessian of the real valued function $F_i$. Notice that if $x$ (and consequently $x_s$) is univariate, i.e., if $x\in\Rset$, then $T(x, x_s, m)$ can be written as $\sum_{v=0}^m \frac{F_i^{(v)}(x_s)}{v!}(x-x_s)^v$. This is not possible for multivariate functions, because $(x-x_s)^v$ is not well defined if $x\in\Rset^{n}$ with $n\in\Zset_{>1}$.

Obviously,
 \begin{equation}\label{eq:taylor}
 F_i(x)=T(x, x, p)=T(x, x_s, \infty), \quad \forall p\geq 0,
 \end{equation}
  is the Taylor series expansion of $F_i$ around the sampling point $x_s$, in the set $\cB_{\delta_{x_s}}(x_s)\subseteq \cS\oplus \cB_{\delta_{x_s}}$. \eqref{eq:taylor} can be rewritten
\begin{align}\label{eq:taylor_rem}
 F_i(x)& =T(x, x_s, \infty) \notag\\
 & =T(x, x_s, \infty)+T(x, x_s, m)-T(x, x_s, m) \notag\\
 & =\underbrace{T(x, x_s, m)}_{\text{m--th order Taylor expansion}}+\underbrace{T(x, x_s, \infty)-T(x, x_s, m)}_{\text{remainder}},
 \end{align}
 which is the $m$-th order Taylor series expansion of $F_i$, with remainder. We refer the reader to the Appendix for further processing of \eqref{eq:taylor_rem} to an equality which replaces the infinite number of terms in the remainder with a Lagrange remainder.

The infinite Taylor series can be over-approximated by a first order Taylor expansion and the Lagrange remainder:
\begin{align}\label{eq:eq7}
F_i(x)= & F_i(x_s)+\nabla F_i(x_s)(x-x_s)+L_1(x, x_s, \xi)
\end{align}
where
$$L_1(x, x_s, \xi)=\frac{1}{2}(x-x_s)^T\nabla^2 F_i(x_s+\xi(x-x_s))(x-x_s)$$
is the Lagrange remainder and $\xi\in (0, 1)$.
For any $x_s\in\cS_s$ there exists $b_{x_s}\in\Rset_+$ such that
\begin{align}\label{eq:eq8}
|L_1(x, x_s, \xi)|\leq b_{x_s}, \quad \forall x\in\cB_{\delta_{x_s}}(x_s), \forall \xi\in(0, 1).
\end{align}
It is possible to compute such bounds for a convex set $\cB_{\delta}(x_{\delta})$, as follows:

\begin{proposition}\label{thm:bound}\cite{althoff2008reachability}
The bounds on the absolute values of the Lagrange remainder in \eqref{eq:eq8} for an $x_s\in \cS_s$, can be computed as follows:
\begin{equation}\label{eq:eqq6}
b_{x_s}^i=\frac{1}{2}\tau_{x_s}^T\max_{x\in \cB_{\delta_{x_s}}(x_s), \xi\in (0, 1)}(|\nabla^2 F_i(x_s+\xi(x-x_s))|)\tau_{x_s},
\end{equation}
where $\tau_{x_s}\in\Rset^n$ and $\tau_{x_s}(i)=\max\{|\delta_{x_s}(2i-1)|, |\delta_{x_s}(2i)|\}$.
\end{proposition}
The proof is similar to the proof in \cite{althoff2008reachability}, but the zonotopes therein reduce here to hyper--rectangles.

The term $\max_{x\in \cB_{\delta_{x_s}}(x_s), \xi\in (0, 1)}(|\nabla^2 F_i(x_s+\xi(x-x_s))|)$ in \eqref{eq:eqq6} can be computed via interval arithmetics, see \cite{jaulin2001} or \cite{moore2009introduction}. In Matlab, efficient interval analysis can be performed via INTLAB \cite{Ru99a}.

\vspace{0.3cm}
By \eqref{eq:eq7}, \eqref{eq:eq8} and the triangle inequality we see that
\begin{align}\label{eq:eq9}
|F_i(x)-F_i(x_s)| = & |F_i(x_s)+\nabla F_i(x_s)(x-x_s)+ \notag\\
& + L_1(x, x_s, \xi)-F_i(x_s)| \notag\\
\leq & |\nabla F_i(x_s)(x-x_s)|+|L_1(x, x_s, \xi)| \notag\\
\leq & \|\nabla F_i(x_s)\|\|x-x_s\|+b_{x_s}^i,
\end{align}
for all $x\in\cB_{\delta_{x_s}}(x_s)$. Denote $a_{x_s}^i:=\|\nabla F_i(x_s)\|$ and \eqref{eq:eq9} becomes:
\begin{align}\label{eq:eq10}
|F_i(x)-F_i(x_s)| & \leq a_{x_s}^i\|x-x_s\|+b_{x_s}^i,
\end{align}
for all $x\in\cB_{\delta_{x_s}}(x_s)$.

Therefore, Assumption~\ref{differentiabillity} implies \eqref{eq:eq10}, which is
identical to \eqref{eq3.222}, which means that Assumption~\ref{differentiabillity} implies Assumption~\ref{ficont1}.


To decrease the conservatism of \eqref{eq:eq10}, we can reduce the size of the bound $b_{x_s}$ as follows. For example, we can modify \eqref{eq:eq9} in the following manner:
\begin{align}\label{eq:eqalternative1}
|F_i(x)-F_i(x_s)| = & |\nabla F_i(x_s)(x-x_s)+ L_1(x, x_s, \xi)| \notag\\
\leq & a_{x_s}^i\|x-x_s\|+b_{x_s}^i,
\end{align}
where $b_{x_s}^i=0$ and $a_{x_s}^i:=\|\nabla F_i(x_s)+\frac{1}{2}(x-x_s)^T \nabla^2 F_i(x_s+\xi(x-x_s))\|$ can be computed via interval arithmetics. In this way, the triangle inequality is not used in \eqref{eq:eq9} and the bound may become less conservative. With this approach, \eqref{eq:eqalternative1} may replace \eqref{eq:eq10}.

\begin{remark}
Providing a guarantee that $F(x)< 0$ for all $x\in\cB_{\delta_{x_s}}(x_s)$ based solely on evaluating $F(x)$ in $x_s$, comes with the price of a conservatism, through the term $-\bar{\gamma}(\max | \delta_{x_s}|, x_s)$. However, notice that for $\delta_{x_s}\rightarrow 0$ it follows that $-\bar{\gamma}(\max | \delta_{x_s}|, x_s) \rightarrow 0$, and thus, when $F$ is continuous on $\cS$, the inequality $F(x)<0$ is asymptotically recovered in $F(x_s)<0$, without conservatism.
\end{remark}

We are now ready to proceed to the main considered application of Theorem~\ref{thm:general}, i.e., sampling--based verification of Lyapunov's inequality.
Theorem~\ref{thm:general} may be used to verify other properties, e.g., invariance, see \cite{bobiti2015seattle}.

\section{Sampling--based stability verification}
\label{stability}

\subsection{Discrete--time systems Lyapunov inequality verification}

In general, to verify $\cK\cL$--stability of system \eqref{eq:nondist} on a compact set $\cS$, we choose a candidate Lyapunov function $W$ which satisfies \eqref{eq:thm2acompact} and we verify that the Lyapunov's inequality \eqref{eq:thm2bcompact} holds for all $x\in\cS$. However, $W$ is difficult to choose. Therefore, we select an \textit{arbitrary} function $V$ (i.e., $\eta\in\cK_{\infty}$) satisfying \eqref{eq:fslf} and we iterate $M$ until \eqref{eq:ftlf_ineq} holds. As it will be detailed further in the paper, due to issues at the origin inequality \eqref{eq:ftlf_ineq} can only be verified in an annulus of $\cS$, i.e. $\cA\subset\cS$. While the construction in \eqref{eq:lf} provides then a Lyapunov function on $\cA$, an additional result will be worked out to conclude $\cK\cL$--stability in $\cS$.

The question of verifying Lyapunov's inequality for system \eqref{eq:nondist} on a compact set $\cA\subset\cS$, is posed in a sampling--based framework as follows.

\begin{problem}
\label{dtstabprob}
Fix a candidate function $V$, which satisfies \eqref{eq:fslf}. Formulate the problem of verifying that \eqref{eq:ftlf_ineq} holds for all $x\in \cA\subset\cS$ with $0\notin \cA$ via Theorem~\ref{thm:general}, with some $M\in\Zset_{\geq 1}$ and a sampling $\cA_s$ of the set $\cS$.
\end{problem}

\algnewcommand{\algorithmicgoto}{\textbf{go to}}%
\algnewcommand{\Goto}[1]{\algorithmicgoto~\ref{#1}}%
\renewcommand{\algorithmicrequire}{\textbf{Input:}}
\renewcommand{\algorithmicensure}{\textbf{Output:}}
\begin{algorithm}
\caption{Construct $\cA \subset \cS$ such that $F(x)<0$ on $\cA$.}\label{lyapver}
\begin{algorithmic}[1]
\Require $\cS$, $V$, $G$, $M$, $\delta_{min}$
\Ensure $\cA$, ($F(x)<0$ on $\cA$)
\Statex
\State $wrong\gets []; r\gets 0; \cA=\varnothing; good\gets []; p\gets 0$
\State Select a finite set of samples $\cA_s\subset\cS$
\ForAll{$x_s\in\cA_s $}
        \State $\bar{\gamma}(\max |\delta_{x_s}|, x_s)=a_{x_s}\max |\delta_{x_s}|+b_{x_s}+\varepsilon(x_s)$
        \If{$F(x_s)>-\bar{\gamma}(\max |\delta_{x_s}|, x_s)$}
            \State $r\gets r+1$
            \State {$wrong(r).del\gets \delta_{x_s}$}
            \State {$wrong(r).spoint\gets x_s$}
            \State {$wrong(r).tau\gets \tau_{x_s}$}
        \Else
            \State $p\gets p+1$
            \State {$good(p).del\gets \delta_{x_s}$}
            \State {$good(p).spoint\gets x_s$}
            \State {$good(p).tau\gets \tau_{x_s}$}
            \State $\cA\gets\cA\cup \cB_{\delta_{x_s}}(x_s)$
        \EndIf
\EndFor

\State $k\gets 1$
\If{$r>0$}
        \While{1}
             \If{$ \max\{wrong(k).del\}>\delta_{min}$}
                \State Generate set $\cB_{\delta_{x_s}}^s(wrong(k).spoint)$ of samples by multi--resolution on $\cB_{\delta_{x_s}}(wrong(k).spoint)$
                \ForAll{$x_s\in\cB_{\delta_{x_s}}^s(wrong(k).spoint) $}
                    \State $\delta_{x_s}=wrong(k).del/2$
                    \State Apply steps 4--15
                \EndFor
             \EndIf
             \If{$k==r$}
                \State break
             \EndIf
             \State $k\gets k+1$
        \EndWhile
\EndIf

\end{algorithmic}
\end{algorithm}
\begin{algorithm}
\caption{Verify Lyapunov inequality on $\cA$.}\label{fullalg}
\begin{algorithmic}[1]
\Require $G$, $\cS$, $\delta_{min}$, $M_{max}$, $V$, $M$
\Ensure $W$, $\cA$
\Statex
\State Algorithm~\ref{lyapver}: Verify that $F(x)=V(G^M(x))-\rho(V(x))<0$ for all $x\in\cA\subseteq \cS$, with minimum resolution $\delta_{min}$. \label{mark}
\If {Algorithm~\ref{lyapver} halts}
    \State $M=M+1$;
    \If {$M<M_{max}$}
        \State \Goto{mark}
    \Else
        \State break;
        \State Hint: select another function $V$
    \EndIf
\Else
    \State $W(x)=\sum_{i=0}^{M-1} V(G^i(x))$
\EndIf
\end{algorithmic}
\end{algorithm}
To approach Problem~\ref{dtstabprob}, express the property function $F(x)$ as follows:
\begin{equation}\label{eq:lyapfun}
F(x):=V(G^M(x))-\rho\left(V(x)\right), \quad \forall x\in\cS,
\end{equation}
with $\rho\in\cK$ which satisfies $\rho<id$.

Algorithm~\ref{lyapver} reports all the operations necessary for verifying that $F(x)<0$ on $\cA \subset\cS$, via Theorem~\ref{thm:general}. As detailed therein, the verification starts from the complete set $\cS$ and gradually the set $\cA$ is constructed by the balls $\cB_{\delta_{x_s}}(x_s)$ which do satisfy $F(x_s)<-\bar{\gamma}(\max |\delta_{x_s}|, x_s)$. Note that Algorithm~\ref{lyapver} illustrates a multi--resolution sampling approach to Theorem~\ref{thm:general} and offers a solution to Problem~\ref{dtstabprob}. To obtain the true Lyapunov function $W$, Algorithm~\ref{fullalg} is executed, which embeds Algorithm~\ref{lyapver}.

Algorithm~\ref{fullalg} starts with a given $M$ and verifies $F(x)<0$ on a set $\cA$ via Algorithm~\ref{lyapver}. If the verification does not provide a set $\cA$, then $M$ is increased until a satisfiable set $\cA$ is achieved. If a maximum $M_{max}$ is reached, it is recommended to choose another function $V$, which may provide a smaller satisfying $M$. The output of Algorithm~\ref{fullalg} is a Lyapunov function $W$, valid on $\cA$.

\begin{remark}
The main step which raises scalability challenges in Algorithm~\ref{fullalg} is step 4, i.e., applying Algorithm~\ref{lyapver}, because of the number of samples, which is an exponential function of the system dimension, and the level of multi--resolution. It is therefore beneficial to exploit the decentralized feature of this algorithm in each sampling point via parallelization. To assess the computational load of Algorithm~\ref{lyapver}, let us assume that the computational cost of computing step 5 for one sampling point is $c\in\Rset_+$. Also, assume that there exists a number $p\in\Zset_+$ of processors and the level of multi--resolution that we employ is $m\in\Zset_+$. Also, denote by $w_i\in\Zset_+$ the number of samples that were not verified at the previous multi--resolution step, where $i\in\Zset_{[1, m]}$. Notice that $w_1$ is the initial number of samples, which is the number of elements in $\cA_s$, at step 2 of Algorithm~\ref{lyapver}. Considering also that by multi--resolution of one hyper--rectangle we obtain $2^{n_r}$ new samples, then, the computational complexity of the \verb"for" loop at steps 3-11 in Algorithm~\ref{lyapver} is of the order
$C=c\left(\ceil{w_1/p}+\ceil{w_2 2^{n_r}/p}+\ldots+\ceil{w_m 2^{n_r}/p}\right).$
Notice that, if the number of processors is unlimited, i.e., $p\rightarrow\infty$, then $C=c*m$, because at every level of multi--resolution, the number of processors in use is the same as the number of sampling points which we verify.
\end{remark}

 If $F$ is continuous in $0$ and if the compact set $\cS$ satisfies $0\in int(\cS)$, as it is the case when we want to verify stability of the origin, then $F(0)=0$, and therefore the inequality $F(x_s)\leq -\bar{\gamma}(\max |\delta_{x_s}|, x_s)$ can not be satisfied for $x_s=0$. Hence, the closer $x_s$  will be to zero, the more conservative the condition becomes. For this reason, the set $\cA$, computed via Algorithm~\ref{lyapver}, is an annulus. To cover the neighborhood around the origin, in this paper we make use of a set $\cL$, which is the level set of a true local Lyapunov function $V_L$. The methodology used here to compute a local Lyapunov function relies on linearization of the dynamics in \eqref{eq:nondist}. If the system is stable in the origin, then, the Lyapunov function found for the linear system is also a Lyapunov function for the nonlinear system in a neighborhood $\cN_1(0)$, see \cite[Theorem 4.7, pag. 139]{khalil2002nonlinear}. The set $\cL$ is then the largest level set of the Lyapunov function $V_L$ inside $\cN_1(0)$.
  In what follows, we propose a method to prove that stability can be guaranteed on $\cA\cup\cL$ with the ingredients we have so far. Before this method is introduced, let us state the following fact.
\begin{fact}\label{thm:invfact}
Let $W$ be a candidate Lyapunov function satisfying \eqref{eq:thm2acompact}. Moreover, $\Wset:=\{x | W(x)\leq L\}$ is a level set of $W$ with $L\in\Rset_{>0}$ and $\cL\subseteq\Wset$ is a compact invariant set for system \eqref{eq:nondist}, with $0\in int(\cL)$. If $W(G(x))-\rho(W(x))< 0$ holds for all $x\in\overline{\Wset\setminus \cL}$ with $\rho<id$, then $\Wset$ is an invariant set.
\end{fact}
\begin{IEEEproof}
The set $\Wset$ is invariant if and only if for all $x\in\Wset$ if holds that $G(x)\in\Wset$. If $x\in\cL$, then $G(x)\in\cL\subseteq\Wset$ by the invariance of $\cL$. Otherwise, if $x\in\overline{\Wset\setminus \cL}$, then $W(G(x))< \rho\left(W(x)\right)\leq \rho(L)<L$, and thus, $G(x)\in int(\Wset)\subset\Wset$, which completes the proof.
\end{IEEEproof}

If the Lyapunov inequality holds on $\cS$, then the system \eqref{eq:nondist} is $\cK\cL$--stable in $\cS$. However, as pointed above, the Lyapunov inequality can not be verified via the sampling--based method at $x_s=0$. Therefore, we can at most verify the Lyapunov inequality via sampling on an annulus $\cA$. Still, with the aid of the next theorem, we can establish $\cK\cL$--stability in $\cS$.

\begin{figure}[!htpb]
      \centering
      \vspace{0.2cm}
      \begin{tikzpicture}
    \draw[thick] (-0.5,-0.5) -- (0.5,-0.5);
    \draw[thick] (-0.5,0.5) -- (-0.5,-0.5);
    \draw[thick] (0.5,0.5) -- (0.5,-0.5);
    \draw[thick] (-0.5,0.5) -- (-0.2,0.5);
    \draw[thick] (0.5,0.5) -- (0.2,0.5);
    \draw[thick] (-0.2,0.7) -- (-0.2,0.5);
    \draw[thick] (0.2,0.7) -- (0.2,0.5);
    \draw[thick] (-0.2,0.7) -- (0.2,0.7);
    \draw[black!30!green] (0,0) ellipse (2.5cm and 1.5cm);
    
    \draw[black!10!blue] (-0.8,-0.8) -- (0.8,-0.8);
    \draw[black!10!blue] (0.8,-0.8) -- (0.8,0.8);
    \draw[black!10!blue] (0.8,0.8) -- (-0.8,0.8);
    \draw[black!10!blue] (-0.8,0.8) -- (-0.8,-0.8);

    \draw[black!30!green,fill=black!30!green] (0,0) circle (0.5ex);
    \draw[red] (0,0) circle (5ex);

    \node[draw=none, black!30!green] at (0,-0.3) {0};
    \node[draw=none] at (0,0.3) {$\mathcal{N}_2(0)$};
    \node[draw=none, red] at (-1,0) {$\mathcal{L}$};
    \node[draw=none, black!10!blue] at (0,1) {$\mathcal{N}_1(0)$};
    \node[draw=none, black!30!green] at (-2,0) {$\mathbb{W}$};
\end{tikzpicture}
      \caption{Set inclusions for Theorem~\ref{thm:joint_simple}.}
      \label{fig3}
\end{figure}
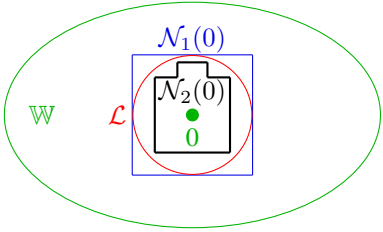

\begin{theorem}\label{thm:joint_simple}
  Let $W$ be a candidate Lyapunov function satisfying \eqref{eq:thm2acompact}, with $\Wset\in\Rset^n$ a level set of $W$. Consider a proper set $\cN_2(0)\subseteq\Wset$. Denote by $\cA_{\Wset}:=\overline{\Wset \setminus \cN_2(0)}$ the annulus of the compact set $\Wset$ with respect to the set $\cN_2(0)$ and suppose that $W(G(x))-\rho(W(x))\leq 0$ holds for all $x\in\cA_{\Wset}$, where $G$ is the nonlinear map of system \eqref{eq:nondist}. Assume that there exists a compact set $\cL$ (see Fig.~\ref{fig3}) with $0\in int(\cL)$ and $\cL\subseteq\Wset$ which is invariant with respect to the nonlinear system \eqref{eq:nondist} and admits a Lyapunov function $V_L:\Rset^n\rightarrow\Rset_+$. Then, system \eqref{eq:nondist} is $\cK\cL$--stable on $\Wset$.
\end{theorem}

\begin{IEEEproof}
Given an arbitrary initial condition $x_0\in\Wset$, the following situations can be encountered:
\begin{enumerate}
  \item If $\overline{x}_0\in \cL$, then Proposition~\ref{thm:2suf_compact} may be applied for system \eqref{eq:nondist} with $W:=V_L$, and therefore system \eqref{eq:nondist} is $\cK\cL$--stable on $\cL$.
  \item If $x_0\in \Wset\setminus\cL\subseteq \cA_{\Wset}$, and since by Fact~\ref{thm:invfact} $\Wset$ is an invariant set, then $G(x_0)\in\Wset$, which allows for two situations:
      \begin{enumerate}
        \item If $G(x_0)\in\cL$, then the reasoning used in case 1). can be applied again with $\overline{x}_0:=G(x_0)$.
        \item If $G(x_0)\in\Wset\setminus\cL\subseteq \cA_{\Wset}$, suppose $\nexists i\in\Zset_{>0}$ such that $G^{i}(x_0)\in\cL$. Then, by the invariance of the set $\Wset$ it follows that $G^{i}(x_0)\in\Wset\setminus\cL\subseteq \cA_{\Wset}$, $\forall i\in\Zset_{>0}$. Therefore, the inequality $W(G(x))-\rho(W(x))\leq 0$ can be iterated $i$--times to obtain the following:
            $$0\leq W(G^{i}(x_0))\leq \rho^i (W(x_0)), \forall i\in\Zset_{>0}.$$
            Because $\rho<id$ and $\rho(0)= 0$, then $\lim_{i\rightarrow\infty} \rho^i(W(x_0))=0$, and therefore
            \begin{equation}\label{eq:ecuatie1}
            \lim_{i\rightarrow\infty} W(G^{i}(x_0))=0.
            \end{equation}
            Moreover, by \eqref{eq:thm2acompact} we know that
            \begin{equation}\label{eq:ecuatie2}
            0\leq\alpha_1(\|G^{i}(x_0)\|)\leq W(G^{i}(x_0)).
            \end{equation}
            By \eqref{eq:ecuatie1} and \eqref{eq:ecuatie2} the following limit holds:
            \begin{equation}\label{eq:ecuatie3}
            \lim_{i\rightarrow\infty} \alpha_1(\|G^{i}(x_0)\|)=0,
            \end{equation}
            which, by the definition of $\cK$--functions, yields:
            \begin{equation}\label{eq:norm1}
            \lim_{i\rightarrow\infty} \|G^{i}(x_0)\|=0.
            \end{equation}
            However, because $G^{i}(x_0)\notin \cL, \forall i\in\Zset_{>0}$, then
            \begin{equation}\label{eq:norm2}
            \|G^{i}(x_0)\|> r_B>0, \forall i\in\Zset_{>0},
            \end{equation}
            and therefore \eqref{eq:norm2} contradicts \eqref{eq:norm1}. This means that $\exists i\in\Zset_{>0}$ such that $G^{i}(x_0)\in\cL$. Thus, the reasoning in case 1) can be applied with $\overline{x}_0:=G^i(x_0)$.
      \end{enumerate}
\end{enumerate}
The above cases cover all the possible trajectory situations starting from the set $\Wset$, and therefore prove $\cK\cL$--stability of system \eqref{eq:nondist} on $\Wset$.
\end{IEEEproof}

The proof follows the same principles as in \cite[Theorem 4.3]{bobiti2015seattle}, while Fact~\ref{thm:invfact} eliminates the requirement of verifying the finite--step invariance of the set $\Wset$. In the case when even a local Lyapunov function $V_L$ can not be found, the safety of the trajectories starting in $\Wset$ can still be guaranteed if
 $$Reach(\cN_2(0))\subseteq \cA\cup\cN_2(0)\subseteq\Wset,$$
 in which case $\Wset$ is guaranteed to be an invariant set.

 This subsection has illustrated a sampling--based verification of the Lyapunov's inequality on $\cA\cup\cL$. The next section shows a method of computing the largest level set $\Wset$ of the Lyapunov function $W$ inside $\cA\cup\cL$. On this set $\Wset$ system \eqref{eq:nondist} is $\cK\cL$--stable, and therefore, $\Wset$ is a subset and an approximation of the DOA of the origin.


\subsection{Level set computation in a sampling--based framework}
\label{level_sec}

Since the set $\cA\cup\cL$ constructed in Algorithm~\ref{lyapver} might be non--convex, and due the complexity of the construction of $\cA$, i.e., $\cA$ is the union of a number of hyper--rectangles centered at the sampling points, a method is required to compute the largest level set $\Wset^*:=\{x: W(x)\leq L^*\}$ of the Lyapunov function $W$ included in $\cA\cup\cL$. However, $\Wset^*$ is difficult to compute, mostly because of the non--convexity of $\cA\cup\cL$. In this paper, we propose computing an estimation of $L^*$ by a value $\overline{L}$, via sampling. This method assumes that the following necessary condition is satisfied:
$$\exists L>0: \cL\subseteq \{x: W(x)\leq L\}\subseteq \cA\cup\cL,$$
and it relies on computing two estimates, $\overline{L}_1$ and $\overline{L}_2$. The first estimate, $\overline{L}_1$, is an estimate of the largest levelset of $W$ which does not intersect the balls $\cB_{\delta_{x_{su}}}(x_{su})$, where $x_{su}$ are  \textit{wrong} points which did not satisfy the sampling--based inequality in Algorithm~\ref{lyapver} at the end of the multi--resolution process. We can not conclude that $\overline{L}_1$ provides the optimum level set $\Wset$, because the set $\Wset_1:=\{x: W(x)\leq \overline{L}_1\}$ might exceed the boundary of the set $\cS$. Therefore, the points on $\partial\cS$ which did satisfy the Lyapunov inequality have to be verified as well. Thus, $\overline{L}_2$, estimates the largest levelset of $W$ which is bounded by $\partial\cA\cap\partial\cS$. Then, an estimation of $L^*$ is given by
$\overline{L}=\min\{\overline{L}_1, \overline{L}_2\}.$

 For computing $\overline{L}_1$ and $\overline{L}_2$, the following steps are required:
 \begin{enumerate}
 \item Select a set of samples $\cS_{su}$, respectively $\partial\cS_s$, via Algorithm~\ref{select}.
 \item For each sample $x_s\in\cS_{su}$, respectively $x_s\in\partial\cS_s$, the minimum level set of $W$ intersecting $\cB_{\delta_{x_{s}}}(x_{s})$ is $\Wset^*_{x_{s}}:=\{x: x\in\cB_{\delta_{x_{s}}}(x_{s}), W(x)\leq L^*_{x_s}\}$, where:
\begin{align}\label{op} L^*_{x_s} = & \min_{x, c} c  \nonumber \\
 \mbox{s.t. } & x\in\cB_{\delta_{x_{s}}}(x_{s}), \\
 \mbox{    } &  W(x)=c. \nonumber \end{align}
 However, to find $L^*_{x_s}$ as in \eqref{op}, an optimization problem has to be solved, which might not be practical. For this reason we approximate $L^*_{x_s}$ with a value $\overline{L}_{x_s}$, via interval analysis.
 \item $\overline{L}_1=\min_{x_{s}\in\cS_{su}} \overline{L}_{x_s},$ $\overline{L}_2=\min_{x_{s}\in\cS_{sb}} \overline{L}_{x_s}.$
 \end{enumerate}

\begin{algorithm}
\caption{Sample points which are to be verified for the computation of $\Wset$.}\label{select}
\begin{algorithmic}[1]
\Require $wrong$, $good$, $\delta$
\Ensure $\cS_{sl}$
\Statex
\State $\cS_{sl}\leftarrow\varnothing$,
\ForAll{i=1:length(wrong)}
    \If {$wrong(i).del\leq \delta$}
    \ForAll{j=1:length(good)}
    \State $del\leftarrow|wrong(i).spoint-good(j).spoint|$
        \If {$del\leq \tau_{wrong(i).spoint}+\tau_{good(j).spoint}$}
            \State $\cS_{sl}\leftarrow \cS_{sl}\cup\{wrong(i).spoint\}$
        \EndIf
    \EndFor
    \EndIf
\EndFor

\end{algorithmic}
\end{algorithm}

\begin{remark}
For step 1), we use Algorithm~\ref{select} as follows. The set $\cS_{su}$ is the set of all the sampling points $x_{s}$ which did not satisfy the sampling--based inequality and which satisfy $\cB_{\delta_{x_{s}}}(x_{s})\cap\cA\neq\varnothing$ and $x_{s}\notin \cL$. This can be achieved by Algorithm~\ref{select}, where the inputs $wrong$ and $good$ are computed as in Algorithm~\ref{lyapver}, $\delta:=\delta_{min}$ and the output is the set $\cS_{su}$.  To compute $\partial\cS_s$, select $n-1$ dimensional balls (by eliminating the hyperplane on which the current sampling point lies) $\cB_{\delta_{x_{s}}}(x_{s})$ which satisfy $\partial\cA\cap\partial\cS=\cup_{x_{s}\in \partial\cS_s} \cB_{\delta_{x_{s}}}(x_{s})$. Once we select a fine sampling $\partial\cS_{sfull}$ of the set $\partial\cS$, we refine $\partial\cS_{sfull}$ as in Algorithm~\ref{select}, where the samples in $\partial\cS_{sfull}$ define the vector $wrong$. The $good$ vector is as in Algorithm~\ref{lyapver}, $\delta$ is $\delta_{x_{s}}$. The output is $\partial\cS_s\leftarrow \partial\cS_{sfull}$.
\end{remark}

\begin{remark}
For step 2), set $F(x):=W(x)$ and find as in \eqref{eq3.2} the parameters $a_{x_{s}}$ and $b_{x_{s}}$ such that
\begin{equation}\label{eq:lev}
|W(x)-W(x_{s})|\leq a_{x_{s}}\|x-x_{s}\|+b_{x_{s}}
\end{equation}
for all $x\in\cB_{\delta_{x_{s}}}(x_{s})$. For all $x^*_{s} \in \Wset^*_{x_{s}}$, the inequality in \eqref{eq:lev} becomes
$W(x_{s})-W(x^*_{s})\leq a_{x_{s}}\max(|\delta_{x_{s}}|)+b_{x_{s}},$
which implies that
$W(x^*_{s})=L^*_{x_s}\geq W(x_{s})-a_{x_{s}}\max(|\delta_{x_{s}}|)-b_{x_{s}}.$
Denote $\overline{L}_{x_s}:=W(x_{s})-a_{x_{s}}\max(|\delta_{x_{s}}|)-b_{x_{s}}$.
\end{remark}


If $\overline{L}>0$, the set $\Wset:=\{x: W(x)\leq \overline{L}\}$
is subset of the DOA of the origin for system \eqref{eq:nondist}. If $\delta_{min}$ and $\delta_{x_{sb}}$ are small enough, it is expected that  $\overline{L}$ is an accurate approximation of $L^*$.

\subsection{Continuous--time systems stability verification}

 If we want to verify $\cK\cL$--stability of the continuous--time system \eqref{eq:nondist_ct} on a set $\Wset_c$ which satisfies $0\in int(\Wset_c)$, via a Lyapunov function candidate $W$, then the verification of \eqref{eq:thm2bcompact_ct} on $\Wset_c\setminus \{0\}$ is required. A sampling--based continuous--time system stability analysis problem can be posed as follows:

\begin{problem}
\label{contver}
 Consider the search space $\cS$ and the discretization of the continuous--time system in \eqref{eq:nondist_ct} to be \eqref{eq:nondist}. Suppose that \eqref{eq:nondist} admits a Lyapunov function $W$, e.g., as found via Algorithm~\ref{fullalg}. Verify whether $W$ satisfies the Lyapunov inequality for \eqref{eq:nondist_ct} on a set $\cA_c\subset\cS$ via Theorem~\ref{thm:general} and find a subset $\Wset_c\subseteq\cS$ of the DOA also for the original continuous--time system \eqref{eq:nondist_ct}.
\end{problem}

To solve Problem~\ref{dtstabprob}, express the property function $F(x)$ as follows:
\begin{equation}\label{eq:lyapfun}
F(x):=\dot{W}(x), \quad \forall x\in\cS,
\end{equation}

The steps required to solve Problem~\ref{contver} are provided in Algorithm~\ref{fullalg_ct}. Firstly, Algorithm~\ref{fullalg} is employed to find a Lyapunov function $W$ and a contractive set $\Wset$ for the discretized system \eqref{eq:nondist}. Then, in order to confirm that $W$ satisfies the Lyapunov inequality for system \eqref{eq:nondist_ct} as well, verify that $F(x)=\dot{W}(x)<0$ for all $x\in\cA_c\subset \cS$ via Algorithm~\ref{lyapver}, for system \eqref{eq:nondist_ct}.

\begin{algorithm}
\caption{Verify the Lyapunov's inequality for continuous--time systems.}\label{fullalg_ct}
\begin{algorithmic}[1]
\Require $G_c$, $\cS$, $\delta_{min}$, $W$
\Ensure $\cA_c$
\Statex
\State Algorithm~\ref{lyapver}: Verify that $F(x)=\dot{W}(x)<0$ for all $x\in\cA_c\subset \cS$, with minimum resolution $\delta_{min}$. \label{mark_ct}
\If {Algorithm~\ref{lyapver} halts}
    \State Select another candidate $W$
    \State go to 1.
\EndIf

\end{algorithmic}
\end{algorithm}

Again, when we want to verify stability of the origin, $F(0)=0$, and therefore the inequality $F(x_s)\leq -\bar{\gamma}(\max |\delta_{x_s}|, x_s)$ can not be satisfied for $x_s=0$. Thus, the set $\cA_c$ is an annulus as well. To cover $\cN_2(0)$, we again upper bound the hole with a set $\cL_c$, which is the level set of a true local Lyapunov function $V_L$. The set $\cL_c$ is then the smallest level set of the Lyapunov function $V_L$ which covers the hole of the annulus $\cA_c$.

 To connect $\cA_c$ with $\cL_c$, we formulate the following result.

\begin{theorem}\label{thm:joint_CT}
Suppose that the compact set $\Wset_c\subset\R^n$ with $0\in int(\Wset_c)$ is defined by the level set $\Wset_c:=\{x | W(x)\leq L_c\}$, where the function $W:\Rset^n\rightarrow\Rset_+$ is continuous and positive definite and $L_c\in\Rset_{>0}$. Consider a proper set $\cB_c\subseteq\Wset_c$. Denote by $\cA_{\Wset_c}:=\overline{\Wset_c \setminus \cB_c}$ the annulus of the compact set $\Wset_c$ with respect to the set $\cB_c$. Assume that there exists a compact set $\cL_c$ with $\cB_c\subseteq\cL_c\subseteq\Wset_c$, which is invariant for system \eqref{eq:nondist_ct}, and admits a Lyapunov function $V_L:\Rset^n\rightarrow\Rset_+$. Moreover, suppose that $\dot{W}(x)< 0$ for all $x\in\cA_{\Wset_c}$. Then, system \eqref{eq:nondist_ct} is $\cK\cL$--stable on $\Wset_c$.
\end{theorem}

\begin{IEEEproof}

Notice that, according to \cite{blanchini2007set}, $\Wset_c$ is a practical set, and therefore, according to Nagumo's theorem, the set $\Wset_c$ is positively invariant w.r.t. \eqref{eq:nondist_ct} if and only if
$$\nabla (W(x)-L)^TG_c(x)=\nabla W(x)^TG_c(x)\leq 0$$
for all $x\in\partial \Wset_c$, where $\partial \Wset_c$ denotes the boundary of the set $\Wset_c$. Since $\partial \Wset_c\subseteq\partial \cA_{\Wset_c}$, and if, according to the hypothesis, it holds that
$$\dot{W}(x)=\nabla W(x)^TG_c(x)< 0, \quad{\forall x\in\cA_{\Wset_c}}.$$
Then it follows directly that
$$\nabla W(x)^TG_c(x)\leq 0, \quad{\forall x\in\partial\Wset_c},$$
and therefore, the fact that $\dot{W}(x)< 0$ for all $x\in\cA_{\Wset_c}$ implies that
$\Wset_c$ is positively invariant for \eqref{eq:nondist_ct}.

Given an arbitrary initial condition $x_0\in\Wset_c$, the following situations can be encountered:
\begin{enumerate}
  \item If $x_0\in \cL_c$, then the problem is solved, because system \eqref{eq:nondist_ct} is $\cK\cL$--stable on $\cL_c$.
  \item If $x_0\in \Wset_c\setminus\cL_c\subseteq \cA_{\Wset_c}$, then, due to the invariance of the set $\Wset_c$ it means that $x(t)\in \Wset_c$, for any $t\in\Rset_+$, which allows for two situations:
      \begin{enumerate}
        \item If $x(t)\in\cL_c$, then the reasoning used in case 1) can be applied again with $\overline{x}_0:=x(t)$.
        \item Assume $x(t)\in\Wset_c\setminus\cL_c\subseteq \cA_{\Wset_c}$ for all $t>0$. By the continuity of $W$ on the compact set $\cA_{\Wset_c}$ the following maximization problem provides a bounded result:
            $$-\beta:=\max_{x\in\cA_{\Wset_c}}\dot{W}(x)<0.$$
            Therefore, for all $x(t)$ starting from an initial condition $x(0)\in\Wset_c\setminus\cL_c\subseteq \cA_{\Wset_c}$ we can write
            \begin{align}
            W(x(t)) & =W(x(0))+\int_0^t\dot{W}(x(\tau))d\tau \nonumber\\
            & \leq W(x(0))-\beta t.
            \end{align}
            Notice that for all $t>\frac{W(x(0))}{\beta}$ it follows that
            $$W(x(t))\leq W(x(0))-\beta t<0,$$
            which contradicts the positive definiteness of $W$. Therefore, for all $x_0\in \Wset_c\setminus\cL_c\subseteq \cA_{\Wset_c}$ there exists $t\in\Rset_{>0}$ such that $x(t)\in \cL_c$, where 2.a) can be applied.
      \end{enumerate}
\end{enumerate}
The above cases prove asymptotic stability of system \eqref{eq:nondist_ct} on set $\Wset_c$.
\end{IEEEproof}

Theorem~\ref{thm:joint_CT} is then employed to conclude that \eqref{eq:nondist} is $\cK\cL$--stable on $\Wset_c$, which can be computed via the procedure in Section~\ref{level_sec}. The set $\Wset_c$ is subset of the DOA of 0 for system \eqref{eq:nondist_ct}. Note that $\Wset_c$ might be different than $\Wset$, which was computed for the discretized system.

\section{Examples}
\label{examples}

This section reveals the potential of the method proposed in this paper for stability analysis and DOA estimation for various systems.

\subsection{2D model}
\label{2D}

This example illustrates the methodology developed in this paper for a 2D discrete--time system.
Consider the system provided in \cite{giesl2007determination}:
\begin{align}\label{eq:sys}
x^+ : = G(x),
\end{align}
 where $x^+$ stands for the next instance of $x$, $x\in\cS\subset\Rset^2$ with $\cS:=\{x | \vert x_1\vert\leq 1, \vert x_2\vert\leq 1.3\},$ and
 $$G(x):=\left[ \begin{array}{c}
        \frac{1}{2}x_1+x_1^2-x_2^2  \\
         -\frac{1}{2}x_2+x_1^2
        \end{array} \right].$$
We will try to find a Lyapunov function for \eqref{eq:sys}, via the steps presented in Algorithm~\ref{fullalg}.

Provide as input in Algorithm~\ref{fullalg} the following: $\delta_{min}=0.02$, $M_{max}=4$. Choose a candidate function $V(x)= x^TPx$ with $P=\left[ \begin{array}{cc}
       10 & 0  \\
        0 & 1
        \end{array} \right]$, which is continuous and two times differentiable, to satisfy Assumption~\ref{differentiabillity}. Fix $M=4$ as a starting value for $M$.
\begin{figure}[!htpb]
      \centering
      \includegraphics[width=0.7\columnwidth]{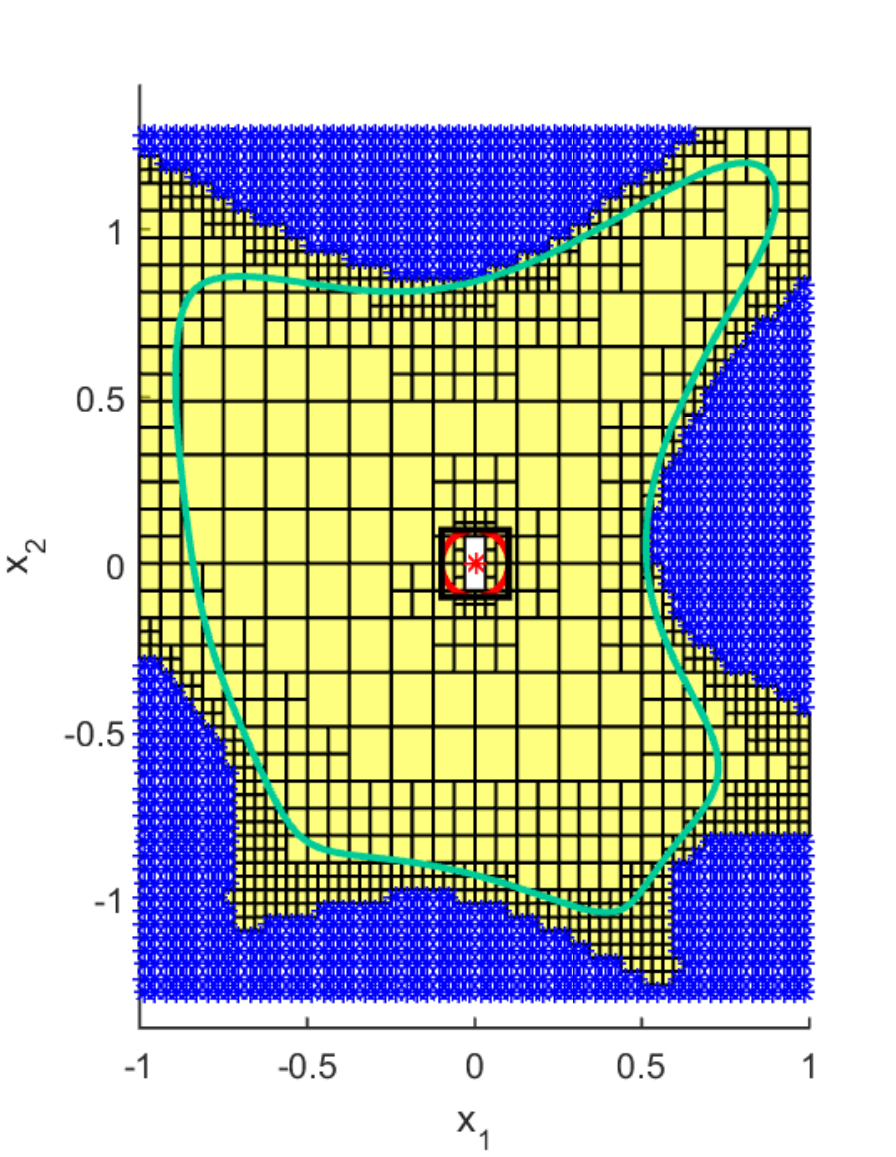}
      \caption{DOA for the origin of the 2D system.}
      \label{fig4}
\end{figure}

For step 2 in Algorithm~\ref{lyapver} we choose $\cA_s=\{0\}$. We sample the set $\cS$ by hyper--rectangles such that initially $\cB_{\delta}(0)=\cS$, where $\delta=[\begin{matrix} 1 & -1 & 1.3 & -1.3\end{matrix}]^T$. We obtain the following: $M=4$, $\cA$ is the yellow set illustrated in Fig.~\ref{fig4}. The white set around the origin is $\cN_2(0)$.

Fix the neighborhood $\cN_1(0)=\{x | \vert x\vert \leq 0.1 \}$. We verify via \texttt{fmincon}, in Matlab, that the quadratic Lyapunov function $V_L$ found for the system linearized in 0 (via \texttt{dlyap}, in Matlab) is also a Lyapunov function for the nonlinear system in $\cN_1(0)$. The maximum level set of the Lyapunov function $V_L(x)=1.3333x^Tx$, of value 0.0133, which is inside $\cN_1(0)$, is an invariant set $\cL$. $\cN_1(0)$ is illustrated with black boundary, $\cL$ with red boundary. The Lyapunov function, as in \eqref{eq:lf}, is
$W(x)=\sum_{i=0}^{3}G^i(x)^TPG^i(x),$
and the set $\Wset$ illustrated with green boundary, i.e., the largest levelset of $W$ (of value $\overline{L}=9.2933$, where $\overline{L}_1=9.2933$ and $\overline{L}_2=11.4642$) which is still contained in $\cA\cup\cL$, estimated according to Section~\ref{level_sec}, is subset of the DOA of the origin.

With blue we have illustrated the points which did not satisfy \eqref{eq:toverify_hybrid} in Theorem~\ref{thm:general} after multi--resolution sampling with $\delta_{min}=0.02$. Simulations of the dynamics starting from these points shows indeed the convergence of most of these trajectories to other equilibria, and not to the origin. Convergence to the maximum $\cA$ can be achieved for $\delta_{min}\rightarrow 0$. Notice in Fig.~\ref{fig4} also the multi--resolution sampling of the set $\cA$. As expected, a more fine resolution is needed towards the boundary of $\cA$, both towards the outer and the inner boundary.

Notice that the set $\Wset$ obtained with the method in this paper is larger than the DOA estimate obtained in \cite{giesl2007determination}. Also, while in \cite{bobiti2016ecc} a total number of $r=3191$ sample points were verified, in this paper, only $r=2012$ sample points were explored, because of the freedom of choosing any rectangle as a sampling unit, which we can maximize in such a manner that we obtain no overlay of the yellow hyper--rectangles, in the construction of $\cA$, as opposed to the unnecessary intersections in \cite{bobiti2016ecc}.

\subsection{Piecewise continuous nonlinear system}
\begin{figure}[!htpb]
      \centering
      \includegraphics[width=1.1\columnwidth]{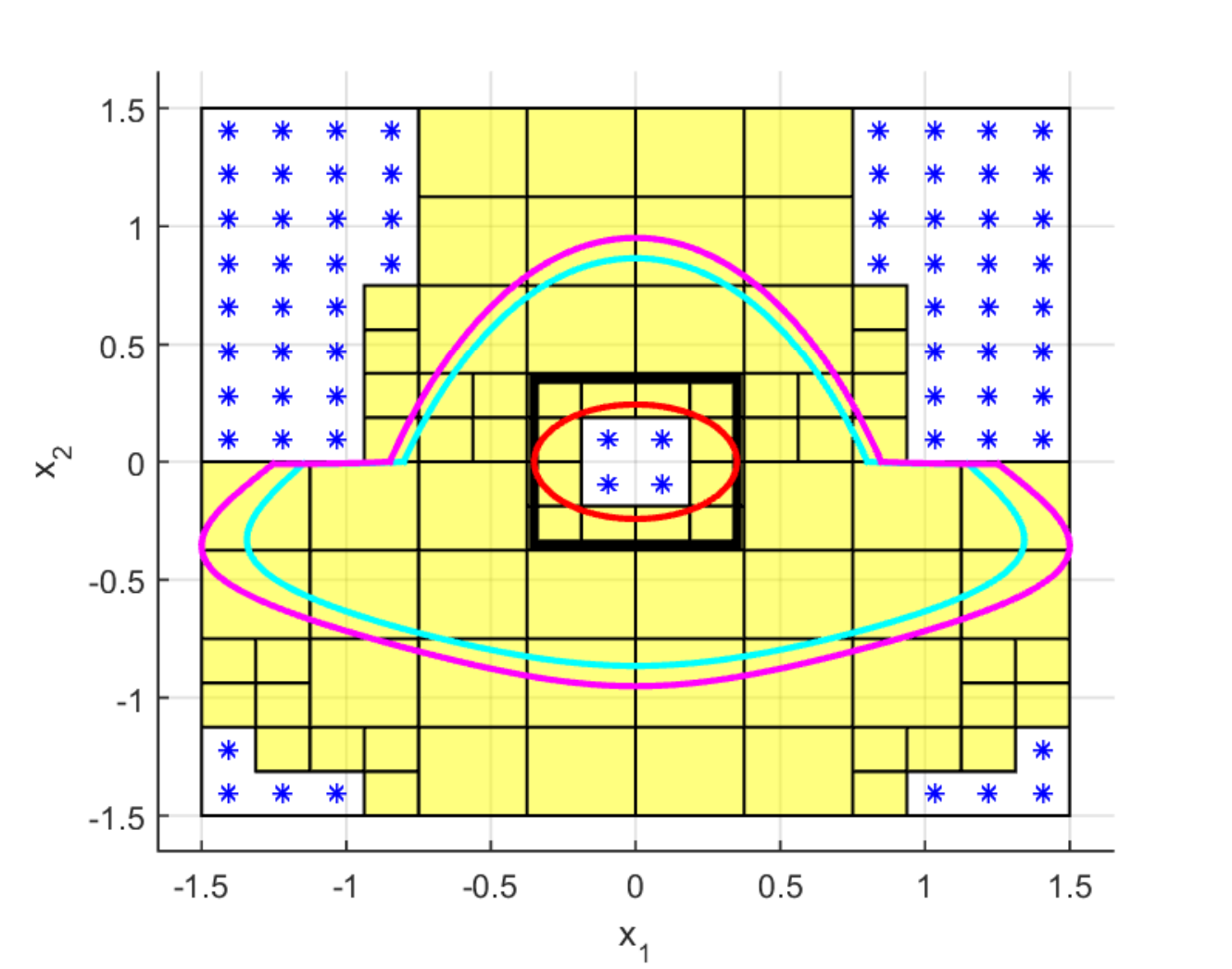}
      \caption{DOA of the origin for the piecewise continuous system.}
      \label{switched}
\end{figure}
To illustrate the method proposed in this paper for verification of a piecewise continuous nonlinear system, consider again Example~\ref{example}. Let the search space be $\cS:=\{x | \vert x_1\vert\leq 1.5, \vert x_2\vert\leq 1.5\}.$

Provide as input in Algorithm~\ref{fullalg} the values $\delta_{min}=0.1$, $M_{max}=3$. Choose a candidate function $V(x)= x^Tx$. By applying Algorithm~\ref{fullalg} to Example~\ref{example}, we obtain the results illustrated in Fig.~\ref{switched}. $\cA$ is the yellow set. With blue we have illustrated the points which did not satisfy \eqref{eq:toverify_hybrid}. The Lyapunov function, found as in \eqref{eq:lf}, is $W(x)=\sum_{i=0}^{3}G^i(x)^TG^i(x).$ The white set around the origin is $\cN_2(0)$.

We linearize $G$ in 0 by linearizing both dynamics $G_1$ and $G_2$ in 0. We obtain a switched linear system, for which $V_L(x)=x^TP_Lx$ with $P_L=\left[\begin{matrix} 26668 & 0 \\ 0 & 55558 \end{matrix}\right]$ is a common Lyapunov function. We choose $\cN_1(0):=\{x | \vert x_1\vert\leq 0.35, \vert x_2\vert\leq 0.35\},$ and via \texttt{fmincon}, in Matlab, we verify that the quadratic Lyapunov function $V_L$ found for the system linearized in 0 is also a Lyapunov function for the nonlinear system in the neighborhood $\cN_1(0)$. The maximum level set of $V_L$ in $\cN$ is $L=3266.8$, which gives the local invariant set $\cL$, illustrated with red.

The set $\Wset$ illustrated with green boundary is the levelset of the Lyapunov function $W$, computed according to Section~\ref{level_sec} and it is subset of the DOA of the origin. Here, $\overline{L}_1=2.3208$, $\overline{L}_2=2.5545$, and therefore $\overline{L}=2.3208$. To compute $\overline{L}_2=2.5545$, we sampled $\partial\cS$ with a distance between samples of value 0.01. Therefore, for a point $x_{su}$ on the vertical boundary of $\partial\cS$, $\delta_{x_{su}}=[\begin{matrix}
0 & 0 & 0.01 & -0.01 \\
\end{matrix}]^T$, while for a point $x_{su}$ on the horizontal boundary of $\partial\cS$, $\delta_{x_{su}}=[\begin{matrix}
0.01 & -0.01 & 0 & 0  \\
\end{matrix}]^T$. $\overline{L}$ provides an underestimation of the true largest levelset of $W$, of value $L^*=2.805$, depicted in Fig.~\ref{switched} with magenta, which is still contained in $\cA\cup\cL$. The relatively low quality of the estimation of the levelset, in this case, is due to the large $\delta_{min}$.

In this example, in step 2 of Algorithm~\ref{lyapver} we choose $\cA_s=\{[\begin{matrix} 0 & 0 \end{matrix}]^T\}$, which is on the switching boundary. The basic sampling unit is a hyper--cube. Note, however, that $\varepsilon([\begin{matrix} 0 & 0 \end{matrix}]^T)=0$. Moreover, by multi--resolution, the samples chosen by the algorithm are not on the switching boundary. Therefore,  $\varepsilon(x_s)=0$ for all $x_s\in\cS_s$. Furthermore, the switching boundaries can always be avoided by choosing a sample outside of the boundary.

In comparison to the results obtained in \cite[Example 4]{luk2015domain}, the DOA computed here is larger in set $S_1$, but smaller in the set $S_2$. Note that in this paper we aim at verifying Lyapunov inequalities and at computing a subset of DOA, without claiming to maximize it.

\subsection{3D model}

The following example illustrates the developed methodology on a 3D system, both in discrete--time, and in continuous--time. The system, see \cite{bjornsson2015class}, is defined by:
\begin{align}\label{eq:sysc}
\dot{x} : = G_c(x)=\left[ \begin{array}{c}
        x_1(x_1^2+x_2^2-1)-x_2(x_3^2+1)  \\
        x_2(x_1^2+x_2^2-1)+x_1(x_3^2+1) \\
        10x_3(x_3^2-1)
        \end{array} \right],
\end{align}
which will be here discretized via the Euler method, i.e., $G(x)=x+hG_c(x)$, where $h=0.1$ is the discretization step.

Let the search space be $\cS:=\{x | \vert x_1\vert\leq 0.9, \vert x_2\vert\leq 0.9, \vert x_3\vert\leq 0.98 \}.$

Consider $\delta_{min}=0.1$ and $M_{max}=2$. Choose a candidate FSLF, $V(x)= x^TPx$ with $P=\left[ \begin{array}{ccc}
       \frac{1}{0.9^2} & 0 & 0  \\
        0 & \frac{1}{0.9^2} & 0 \\
        0 & 0 & \frac{1}{0.98^2}
        \end{array} \right]$. The choice is motivated by the fact that the largest possible ellipsoid that can be contained in the given box $\cS$ is a level set of $V$. Take $M=2$.

We sampled here via hyper--cubes, and we obtain the following: $M=2$, the points in the box $\cS$ which did not satisfy \eqref{eq:toverify_hybrid} in Theorem~\ref{thm:general} are illustrated with blue.

\begin{figure}[!htpb]
      \centering
      \includegraphics[width=1.1\columnwidth]{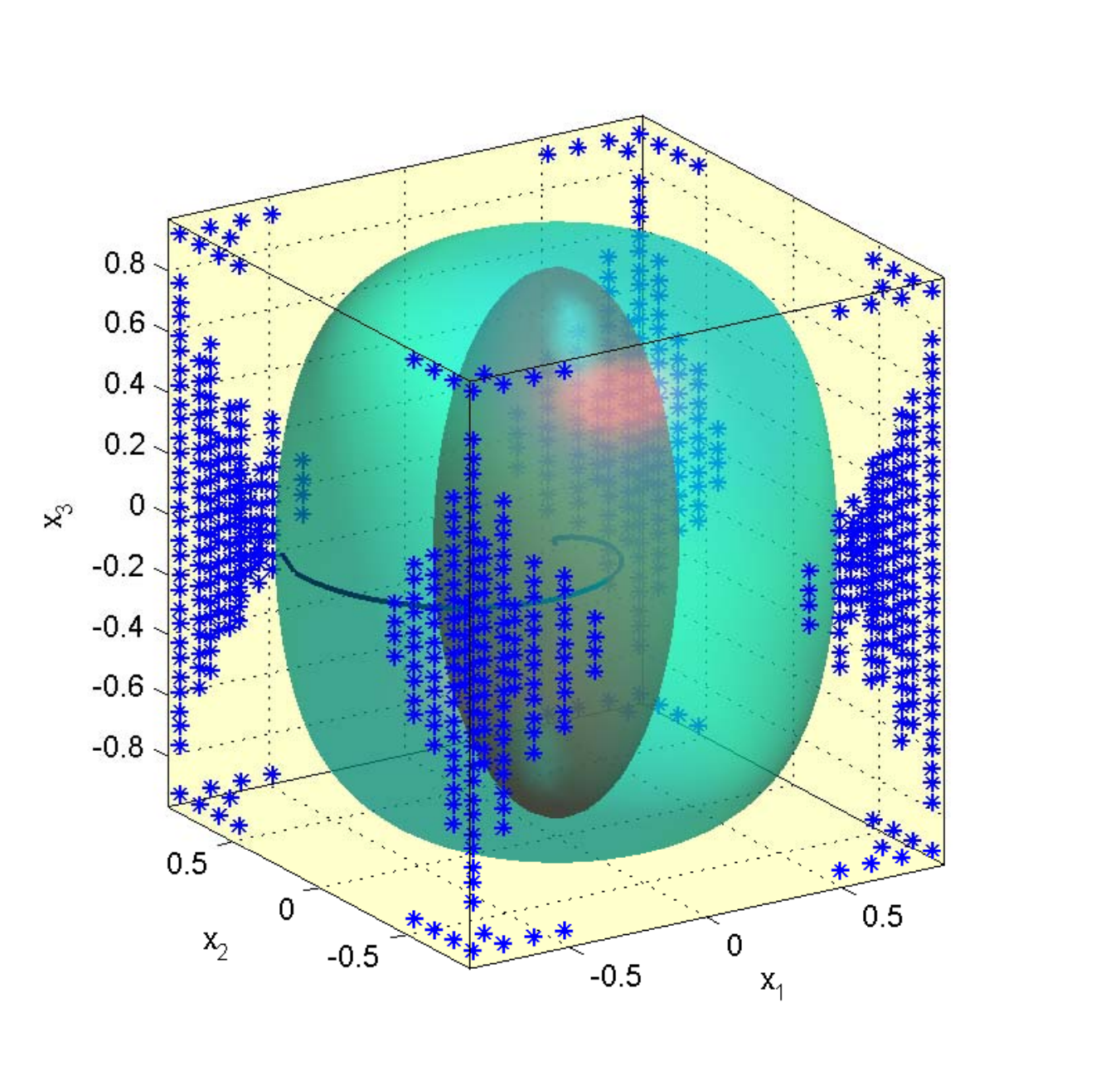}
      \caption{DOA for the origin of the 3D system.}
      \label{fig5}
\end{figure}

Fix the neighborhood $\cN_1(0)=\{x | \vert x_1\vert\leq 0.6, \vert x_2\vert\leq 0.6, \vert x_3\vert\leq 0.9 \}$. Again, we check via \texttt{fmincon}, in Matlab, that the quadratic Lyapunov function
$$V_L(x)=x^T\left[ \begin{array}{ccc}
       5.5556 & 0 & 0  \\
        0 & 5.5556 & 0 \\
        0 & 0 & 1
        \end{array} \right]x$$
found for the system linearized in 0 (via \texttt{dlyap}, in Matlab) is also a Lyapunov function for the nonlinear system in $\cN_1(0)$. The maximum level set of the Lyapunov function $V_L$, of value 0.81, which is inside $\cN_1(0)$, is an invariant set $\cL$, illustrated in Fig.~\ref{fig5} by the red ellipsoid.

The Lyapunov function found with this procedure is $W(x)=\sum_{i=0}^{1}G^i(x)^TPG^i(x).$ The set $\Wset$, illustrated with green, is the largest estimated levelset of $W$ (of value $\overline{L}=1.8459$, where $\overline{L}_1=1.8584$ and $\overline{L}_2=1.8459$) which is still contained in $\cA\cup\cL$. $\Wset$ is subset of the DOA of the origin.

Moreover, by applying Algorithm~\ref{fullalg_ct}, we obtain that $\Wset$ is also subset of the DOA of the original continuous--time system. For the continuous--time system, $\cA_c\neq\cA$, therefore, $\overline{L}_1$ will differ, with a new value $\overline{L}_1=2.0253$. $\overline{L}_2$ has the same value as previously, i.e., $1.8459$, because the set $\cS$ remains the same, and thus $\overline{L}=1.8459$, which means that $\Wset_c=\Wset$.

It is noticeable that the set $\Wset$ contains also regions of the state space which are not found in the DOA computed in \cite{bjornsson2015class} for the original continuous--time system, see, e.g., the black trajectory illustrated in Fig.~\ref{fig5}, having as initial state one of the points which was not captured in \cite{bjornsson2015class}, but which belongs to $\Wset$.

\subsection{Powertrain Control System}

This example illustrates the potential of the methodology developed in this paper for computing the DOA of the origin for a system inspired by a real--life application. Consider a 3D simplified version of a Powertrain Control system, inspired by Example 5 of \cite{kapinski2014simulation}:
\begin{align}\label{eq:sys_PCS}
\dot{x} & : = G_c(x) \nonumber\\
& =\left[ \begin{array}{l}
        c_1\left(2u_1 \sqrt{\frac{p}{c_{11}}-\left(\frac{p}{c_{11}}\right)^2}\right)- \\  \quad\quad-c_1(c_3+c_4c_2p+c_5c_2p^2+c_6c_2^2p)  \\
        4\left(\frac{1}{c_{13}(1+i+c_{14}(r-c_{16}))}-r\right) \\
        c_{15}(r-c_{16})
        \end{array} \right],
\end{align}
where $x=[p\mbox{ }r\mbox{ }i]^T$ is the state vector. Here $p$ is the pressure manifold, $r$ is the air--to--fuel ratio and $i$ is a PI controller, designed to maintain the air--to--fuel ratio in $10\%$ of the optimal value. The corresponding parameters are: $c_1=0.41328$, $c_2=200$, $c_3=-0.366$, $c_4=0.08979$, $c_5=-0.0337$, $c_6=0.0001$,
$u_1=16$,
$c_{11}=1$,
$c_{13}=0.9$,
$c_{14}=0.4$,
$c_{15}=0.4$, $c_{16}=1$. We aim to compute a set $\Wset$ where the control system maintains the performance specification of keeping the air--to--fuel ratio in $10\%$ of the optimal value.
\begin{figure}[!htpb]
      \centering
      \includegraphics[width=1\columnwidth]{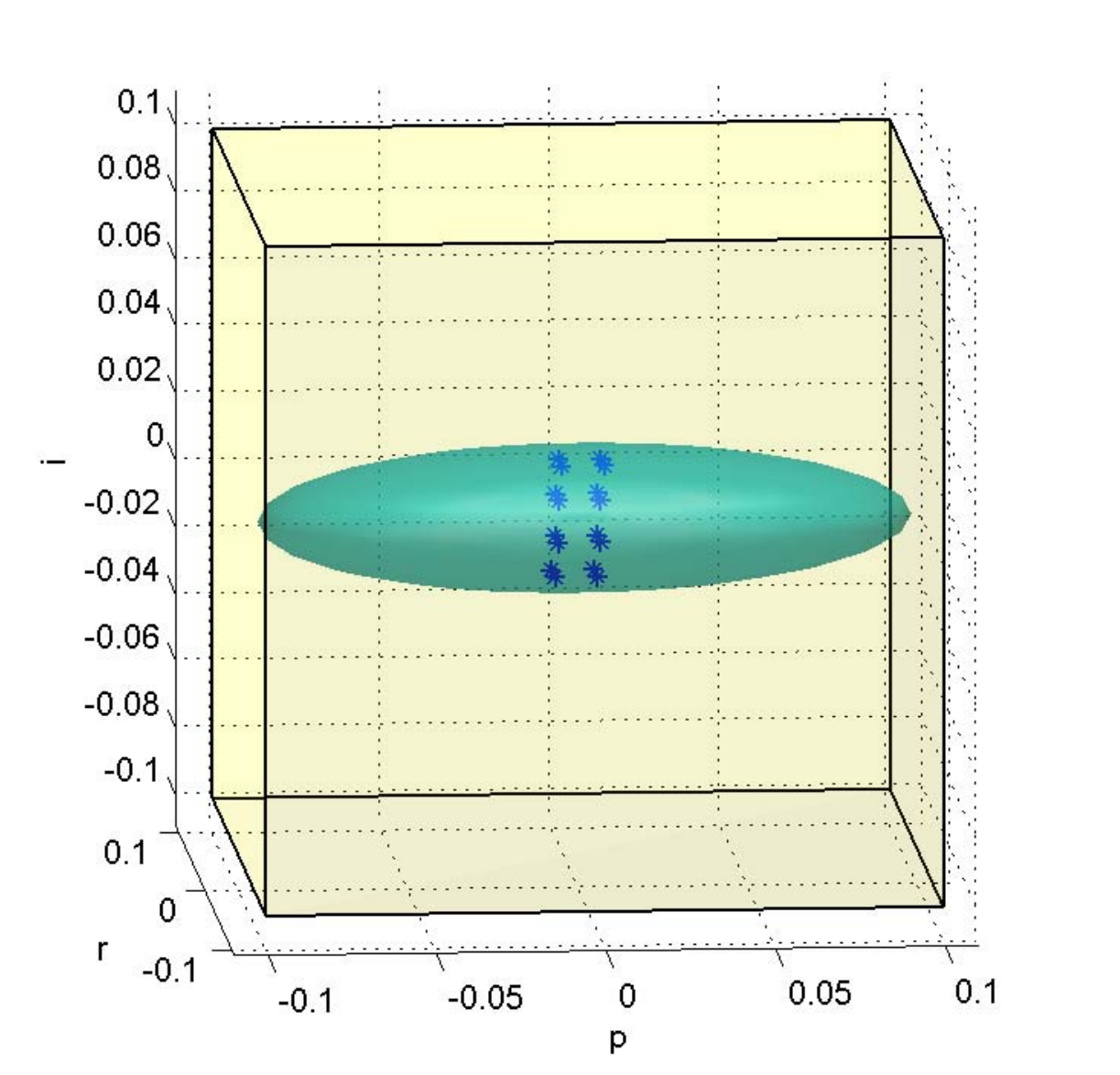}
      \caption{DOA for the origin of the Powertrain Control System.}
      \label{fig6}
\end{figure}
The continuous--time system will be again discretized via the Euler method, i.e., $G(x)=x+hG_c(x)$, where $h=0.01$ is the discretization step. The equilibrium $x_0=[0.7975\mbox{ }1\mbox{ }0.1111]^T$ is translated in 0, and we study the $\cK\cL$--stability of 0, and its corresponding DOA.

Let the search space be $\cS:=\{x | \|x\|\leq 0.1 \}.$ Choose the inputs $\delta_{min}=0.01$ and $M_{max}=3$. Choose a candidate FSLF, $V(x)= x^TPx$ with $P=\left[ \begin{array}{ccc}
       1 & 0 & 0  \\
        0 & 4 & 2 \\
        0 & 2 & 14
        \end{array} \right]$, and $M=3$.

We obtain the following: $M=3$, $\cA$ is the gray set illustrated in Fig.~\ref{fig6}, excluding the balls $\cB_{\delta_{x_s}}(x_s)$, where $x_s$ are the blue points in the box $\cS$ which did not satisfy \eqref{eq:toverify_hybrid} in Theorem~\ref{thm:general}. However, we certify $V(G^3(x))-\rho( V(x))<0$ with $\rho=0.999 id$ by optimization in each of the sets $\cB_{\delta_{x_s}}(x_s)$ via \texttt{fmincon}.

\texttt{fmincon} fails to provide us with certification for a local Lyapunov function $V_L$. Thus we have no invariant set $\cL$. Thus, Theorem~\ref{thm:joint_simple} is not used here, because set $\cS$ is the union of all the sets $\cB_{\delta_{x_s}}(x_s)$ certified via \texttt{fmincon} and the set $\cA$ certified via Theorem~\ref{thm:general}, which implies that $V(G^3(x))-\rho( V(x))<0$ holds for all $x\in\cS$.

The Lyapunov function we find is
$W(x)=\sum_{i=0}^{2}G^i(x)^TPG^i(x),$
and the set $\Wset$ illustrated with green, i.e., the largest levelset of $W$ ($\overline{L}=0.0209$, computed with $\max(|\delta_{x_{sb}}|)=0.02$) which is still contained in $\cS$, is subset of the DOA of the origin for the discretized system. 

By applying Algorithm~\ref{fullalg_ct}, we obtain that $\Wset$ is also subset of the DOA of the original continuous--time system, because $\cA_c=\cA$, which means also that the new levelset of $W$ is the same as computed previously for the discretized system, i.e., $\overline{L}=0.0209$. This fact guarantees that, for any initial condition starting in $\Wset$, $\|r(t)-x_0(2)\|\leq 0.1$, for all $t\in\Rset_+$, which means that, indeed, for any initial condition starting in $\Wset$, air--to--fuel ratio is maintained in $10\%$ of the optimal value.

\section{Conclusions}
\label{conclusions}

In this paper, a sampling--based approach to stability verification for hybrid nonlinear systems via Lyapunov functions was proposed, to avoid large, possibly non--feasible optimization problems involved in finding Lyapunov functions. This constructive approach, applicable to both discrete--time and continuous--time systems, proposes verification of the decrease condition for a candidate Lyapunov function on a finite sampling of a bounded set of initial conditions and then it extends the validity of the Lyapunov function to an infinite set of initial conditions by exploiting continuity properties. Multi--resolution sampling is employed to perform efficient state--space exploration and hyper--rectangles are used as basic sampling blocks, to account for different constraint scale on different states and further reduce the amount of samples to be verified. This verification method is decentralized in the sampling points, which makes the method scalable to any degree. The potential of the proposed methodology was illustrated through examples.





\bibliographystyle{IEEEtran}
\bibliography{Ruxandra}

\section*{APPENDIX}

In what follows we will need to use the Mean Value Theorem to process the Taylor series expansion in \eqref{eq:taylor_rem}, see \cite[Section 2]{berz1998computation}. However, the Mean Value Theorem can not be directly applied to multivariate functions, and therefore a one--dimensional function $f_R:[0, 1]\rightarrow\Rset$ is introduced by the formula $f_R(s)=F_i(x_s+s(x-x_s))$, where $x$ and $x_s$ are given. Note that
$$f_R^{(v)}(s)=([(x-x_s)\nabla]^v F_i)(x_s+s(x-x_s)),$$
and
$$f_R^{(v)}(0)=([(x-x_s)\nabla]^v F_i)(x_s).$$
Denote
$$T_R(s, s_0, m):=\sum_{v=0}^m\frac{f_R^{(v)}(s_0)}{v!}(s-s_0)^v.$$
Notice that the formulae of $T$ and $T_R$ are similar, but $T$ is multivariate, while $T_R$ is univariate. Furthermore
\begin{align}
T_R(1, 0, m) & =\sum_{v=0}^m\frac{f_R^{(v)}(0)}{v!}=\sum_{v=0}^m\frac{([(x-x_s)\nabla]^v F_i)(x_s)}{v!} \notag \\
& = T(x, x_s, m). \notag
\end{align}

Apply Taylor expansion formula to $f_R$ around $0$ and evaluate in $s=1$:
\begin{align}
f_R(1) & =T_R(1, 1,m)=T_R(1, 0,\infty) \notag \\
& =T_R(1, 0,m)+ remainder. \notag
\end{align}
Let us apply the Mean Value Theorem to $T_R$ and an arbitrary function $g:\Rset\rightarrow\Rset$ with $g'(x)\neq 0$ on $(0, 1)$. There exists $\xi\in(0, 1)$ such that:
$$\frac{T_R(1, 1,m)-T_R(1, 0,m)}{g(1)-g(0)}=\frac{T_R'(1, \xi, m)}{g'(\xi)},$$
and therefore
 \begin{equation}\label{eq:meanval}
 T_R(1, 1,m)=T_R(1, 0,m)+\frac{g(1)-g(0)}{g'(\xi)}T_R'(1, \xi, m).
 \end{equation}
To analyse the expression of $T_R'(1, \xi, m)$ when we derivate $T_R$ with respect to the second argument we see that
\begin{align}\label{eq:trder}
T_R'(1, \xi, m) & =\sum_{v=0}^m\left(\frac{f_R^{(v+1)}(\xi)}{v!}(1-\xi)^v-\frac{f_R^{(v)}(\xi)}{v!}(1-\xi)^{v-1}\right) \notag \\
& = \frac{f_R^{(m+1)}(\xi)}{m!}(1-\xi)^m.
\end{align}
If $g(\xi)=(1-\xi)^{m+1}$, then $g'(\xi)=-(m+1)(1-\xi)^{m}$. Notice that $g'(\xi)\neq 0$ for $\xi\in(0,1)$. Then, from \eqref{eq:meanval} and \eqref{eq:trder} it follows that
$$T_R(1, 1,m)=T_R(1, 0,m)+L_m(x, x_s, \xi),$$
where
\begin{align}\label{eq:lagrem}
L_m(x, x_s, \xi)= & \frac{-1}{-(m+1)(1-\xi)^m}(1-\xi)^m\frac{f_R^{(m+1)}(\xi)}{m!} \notag\\
= & \frac{f_R^{(m+1)}(\xi)}{(m+1)!} \notag \\
= & \frac{([(x-x_s)\nabla]^{m+1}F_i)(x_s+\xi(x-x_s))}{(m+1)!}
\end{align}
is the Lagrange remainder and $\xi\in(0, 1)$.

\end{document}